\documentclass[aps,prb,twocolumn,showpacs,amsfonts,amsmath,floatfix,byrevtex]{revtex4-1}

\usepackage{graphics}
\usepackage{graphicx}
\usepackage{amsmath}
\usepackage{longtable}
\usepackage{bm}
\usepackage{epsfig}

\begin{document}


\title{Anomalous magnetization of a carbon nanotube as an excitonic insulator}

\author{Massimo Rontani}

\email{massimo.rontani@nano.cnr.it}
\homepage{www.nano.cnr.it}

\affiliation{CNR-NANO Research Center S3, Via Campi 213a, 41125 Modena, Italy}




\date{\today}

\begin{abstract}
We show theoretically that an undoped carbon nanotube might be 
an excitonic insulator---the long-sought 
phase of matter proposed by Keldysh, Kohn and others fifty years ago.
We predict that the condensation of triplet excitons,
driven by intervalley exchange interaction,
spontaneously occurs at equilibrium if the tube radius is sufficiently small.
The signatures of exciton condensation 
are its sizeable contributions to both the energy gap and 
the magnetic moment per electron. The increase of the gap might
have already been measured, albeit with a different explanation 
[V. V. Deshpande, B. Chandra, R. Caldwell, D. S. Novikov, J. Hone,
and M. Bockrath, Science {\bf 323,} 106 (2009)]. 
The enhancement of the quasiparticle 
magnetic moment is a pair-breaking effect that 
counteracts the weak paramagnetism
of the ground-state condensate of excitons. 
This property could rationalize
the anomalous magnitude of magnetic moments
recently observed in different devices
close to charge neutrality.
\end{abstract}

\pacs{73.63.Fg, 71.35.Lk, 71.35.Ji, 71.70.Gm}

\maketitle

\section{Introduction}

After twenty years of intense investigation, carbon nanotubes
(CNTs) still allow us 
to explore novel quantum physics in one 
dimension.\cite{Dresselhaus1998,Ando2005,McEuen2010,Deshpande2010,Marcus2010,Laird2014}
The quality of suspended tubes\cite{Cao2005}
achieved in transport measurements
has disclosed subtle effects
that previously had been obscured by sample disorder, 
such as spin-orbit interaction\cite{Kuemmeth2008,Churchill2009,Jhang2010,Jespersen2011,Steele2013,Cleuziou2013} 
and Wigner localization.\cite{Deshpande2008,Pecker2013} 
A growing body of experiments on single-wall CNTs shows that  
electron-electron interactions play a prominent role, 
being long-ranged and poorly screened close to charge neutrality. 
These observations include
the emergence of an energy-gap of allegedly many-body origin
in nominally metallic tubes,\cite{Deshpande2009}
Wigner localization of excess charge carriers
in semiconducting tubes,\cite{Deshpande2008,Pecker2013} and the evidence
of strong excitonic effects\cite{Maultzsch2005,Wang2005,Zaric2006,Mortimer2007,Shaver2007,Srivastava2008,Matsunaga2008,Torrens2008}---even in metallic tubes.\cite{Wang2007}   

The standard model of interacting electrons
in CNTs is the Luttinger liquid,\cite{Kane1997b,Giamarchi2004} 
which was successful in explaining 
tunneling\cite{Bockrath1999} and photoemission\cite{Ishii2003} 
spectra of metallic tubes.  
The reason is the perfect mapping of the linearly
dispersive Tomonaga-Luttinger model 
onto the CNT effective-mass Hamiltonian,\cite{Balents1997,Egger1997}  
which at low energies exhibits
the Dirac-Weyl form peculiar to massless fermions.
However, this mapping becomes a poor approximation for undoped semiconducting
tubes,\cite{Levitov2003,Konik2011,Secchi2012}
since in the energy range close to band edges---where interactions 
are most effective---the noninteracting energy spectrum is massive and
the Fermi level undefined. 
This is true even for nominally metallic tubes at half filling,
due to the ubiquitous presence of small mass gaps induced by strain,
twists, curvature,\cite{Kane1997,Charlier2007} 
and spin-orbit coupling.\cite{Ando2000,Huertas-Hernando2006} 
The latter term affects also
armchair tubes, whose metallicity is otherwise protected by 
symmetry.\cite{Kane1997}

Two alternative paradigms of strongly correlated insulators 
might fit carbon nanotubes.
Intriguingly, both models were introduced by Mott 
long ago.\cite{Mott1937,Mott1961,Mott1968}
The first concept is the Mott-Hubbard metal-insulator
transition, which applies to 
solids that are metallic in the absence 
of interactions.\cite{Auerbach1994}
This scenario has been recently put forward\cite{Balents1997,Krotov1997,Nersesyan2003} 
to explain the many-body
energy gap measured in half-filled CNTs.\cite{Deshpande2009} 
This quantity was obtained
after subtracting the contributions to the transport gap 
due to finite-size effects and
noninteracting mechanisms. 
We notice that the
spin-orbit contribution to the energy gap was not considered by 
Ref.~\onlinecite{Deshpande2009}, 
despite the fact that it may be large\cite{Steele2013} 
and cannot be fully compensated by the magnetic field in both spin channels 
(see Sec.~\ref{s:exp}).
 
\begin{figure}
\setlength{\unitlength}{1 cm}
\begin{picture}(8.5,6.0)
\put(0.5,0.0){\epsfig{file=fig1a.eps,width=3.0in,,angle=0}}
\put(3.2,1.3){\epsfig{file=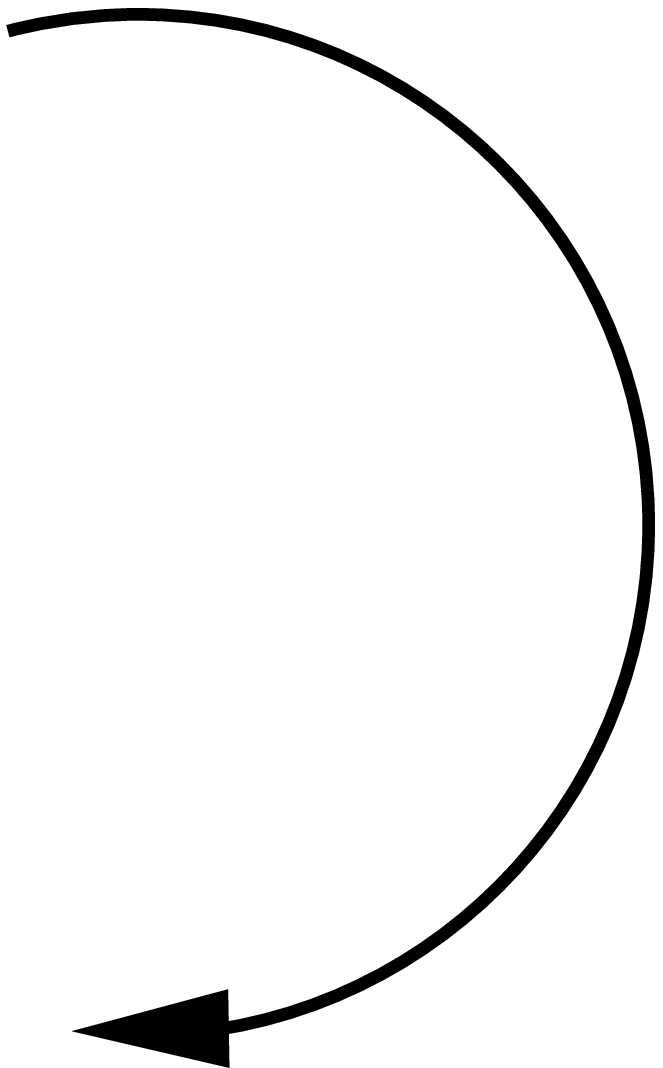,width=0.2in,,angle=0}}
\put(3.2,3.5){\epsfig{file=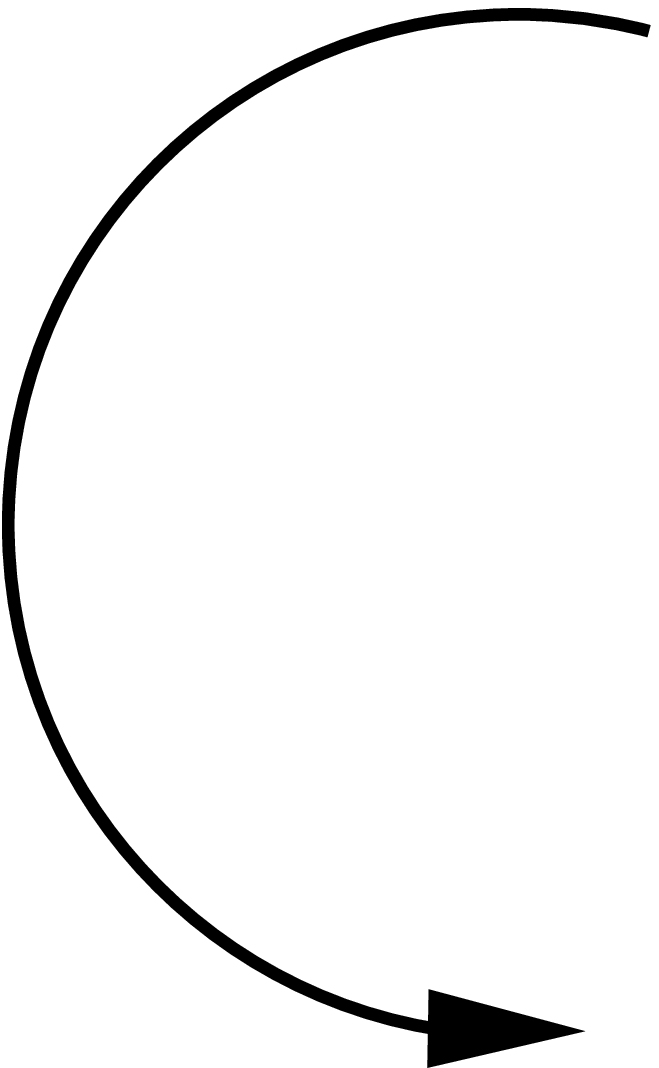,width=0.2in,,angle=0}}
\end{picture}
\caption{(color online)
Orbital magnetization of the exciton condensate.
(a) Normal ground state of a metallic nanotube at finite magnetic field.
The green [light gray] dots point to electrons filling the valence band.
Conduction- and valence-band states exhibit opposite magnetic moments
$\mu_0$, corresponding to either clockwise or anticlockwise rotations of
electrons around the circumferential direction.
The ground state is paramagnetic.
(b) Excitonic insulator (EI) ground state. The magnetization of the EI phase
is reduced with respect to that of the normal phase
due to the spontaneous condensation of excitons. The thick red [gray] lines
are the quasiparticle energy bands renormalized by
the excitonic gap $\Delta$.
\label{f:clock}
}
\end{figure}

The second paradigm is the excitonic insulator (EI)---the focus 
of this work---which applies to semiconductors exhibiting low 
dielectric screening
and nested electron and hole Fermi 
surfaces.\cite{Mott1961,Keldysh1964,Kozlov1965,Kohn1967,Halperin1967,Kohn1968,Volkov1975}  
In the EI the strong attraction between
electrons in the conduction band and holes in the valence band 
leads to the formation of excitons that undergo Bose-Einstein condensation,
similarly to the way Cooper pairs condense in the 
Bardeen-Cooper-Schrieffer (BCS) ground-state.
The outcome is a permanent insulating phase
with energy gap enhanced by the remnant of the exciton
binding energy $\Delta$ (see Fig.~\ref{f:clock}
and Ref.~\onlinecite{Rontani2013} for a recent review).

This scenario has been overlooked so far
despite the strong evidence of major
excitonic effects in CNTs.\cite{Maultzsch2005,Wang2005,Zaric2006,Wang2007,Mortimer2007,Shaver2007,Srivastava2008,Matsunaga2008,Torrens2008} 
One important exception was the
seminal 1997 paper by Ando,\cite{Ando1997} who realized 
that the electron-hole symmetry of CNT
energy bands provides the perfect nesting
for exciton condensation.
However, he (and later Hartmann and coworkers\cite{Hartmann2011})
negated the stability of the EI phase on the basis of the
energetics of spinless excitons.

Here, building on the investigations by 
Ando,\cite{Ando1997,Ando2006} we take a step forward 
and consider `dark' triplet excitons, which have the lowest
energy and are thought to play a prominent role in 
optical experiments.\cite{Zhao2004,Spataru2004,Maultzsch2005,Ando2006,Zaric2006,Mortimer2007,Shaver2007,Srivastava2008,Matsunaga2008,Torrens2008}
Contrary to previous work,\cite{Ando2006,Hartmann2011} 
we treat exchange interactions between electrons and holes in
different valleys in a non-perturbative
manner, as these interactions are poorly screened even for vanishing gap
(the theory by Hartmann {\it et al.}\cite{Hartmann2011} 
does not include the valley degree of freedom).
Hence, including intervalley exchange forces from the beginning
in a model Bethe-Salpeter equation that we solve exactly,
we show that triplet excitons condense 
for sufficiently large values of intervalley exchange interaction
and small radii.
The resulting EI phase differs in many respects 
from that envisioned in the Sixties for semiconductors with parabolic
bands,\cite{Kozlov1965,Kohn1967,Halperin1967,Kohn1968} 
as a consequence of the relativistic dispersion of Dirac fermions and 
tube topology. 

We predict that:
(i) The stability of the EI is independent from the size of the 
noninteracting energy gap,
which may be tuned e.g.~by an axial magnetic field.
(ii) The excitonic gap $\Delta$ adds quadratically to any
noninteracting mass term, including the ubiquitous spin-orbit term. 
(iii) The quasiparticle magnetic moment, as observed
by tunneling spectroscopy,\cite{Minot2004} is enhanced 
with respect to its semiclassical value. 
This latter effect is due to the ionization of one of the excitons
merging the condensate in order 
to release an unbound electron (hole) in the conduction (valence) band.
Such ionization increases the total magnetization with respect to that
of the ground state, which is a weak paramagnet (Fig.~\ref{f:clock}).

\begin{table}
\begin{ruledtabular}
\begin{tabular}{ c c c }
& Mott insulator           
& Excitonic insulator \\      
Transport gap
&  opened
& widened \\
Magnetic moment $\mu$
& unchanged
& enhanced \\
Spin-orbit coupling
& unchanged
& enhanced \\
Subgap excitations 
& yes 
& yes
\end{tabular}
\caption {Mott-Hubbard versus excitonic insulators.
}
\label{t:comparison}
\end{ruledtabular}
\end{table}

The above predictions may be experimentally validated
as well as they may be used to extract the intervalley exchange strength,
which is a fitting parameter in our theory. 
We stress that prediction
(iii), which has no counterpart in the Mott-Hubbard scenario,
might shed light on the anomalous magnitude of magnetic moments
recently reported by different 
groups\cite{JarilloHerrero2005,Kuemmeth2008,Jespersen2011b,Steele2013}
for devices with very few
carriers (cf.~Table \ref{t:comparison} and Sec.~\ref{s:exp}). 
Furthermore, prediction
(ii) could explain the unusually large value of
spin-orbit interaction measured by the Delft group,\cite{Steele2013} 
which 
is presently not understood. Besides, the findings of 
Ref.~\onlinecite{Deshpande2009}
might be
consistent with the EI scenario presented here, including the observation
of subgap neutral excitations, as we further discuss in Sec.~\ref{s:exp}.

The reader not interested in the full derivation of our theory may skip
the more technical sections   
and refer directly to the 
results illustrated in Secs.~\ref{s:instability}, 
\ref{s:anomalous}, \ref{s:weak}, and \ref{s:exp}.
Sections \ref{s:model}, \ref{s:allband}, and \ref{s:intervalley} 
review the results obtained by Ando in Refs.~\onlinecite{Ando1997}
and \onlinecite{Ando2006} as they are the starting point of our development, 
which is presented and discussed in the remaining part of the article.

The structure of this paper is as follows:
We summarize
the effective-mass theory of CNT single-particle states
in Sec.~\ref{s:model}, which was the basis of 
the study of excitons by Ando. In Sec.~\ref{s:allband}
we recall his solution of the Bethe-Salpeter equation for
spinless excitons within the random-phase approximation. 
We use Ando's result to introduce
a simpler two-band model for Dirac excitons 
in Sec.~\ref{s:twoband}. 
We add spin and valley degrees of
freedom in Sec.~\ref{s:intervalley} and solve the corresponding
Bethe-Salpeter equation for the triplet exciton in 
Sec.~\ref{s:BSEtriplet}.
We discuss the resulting excitonic instability 
in Sec.~\ref{s:instability}
whereas in Sec.~\ref{s:ei} we build the many-body 
theory of the EI and solve the gap equation 
in Sec.~\ref{s:gap}.
Then in the following sections we present our main results: 
the anomalous enhancement
of the magnetic moment per particle (Sec.~\ref{s:anomalous}), the 
weak paramagnetism of the EI ground state (Sec.~\ref{s:weak}),
and the relation to experiments 
(Sec.~\ref{s:exp}). After the conclusions (Sec.~\ref{s:end}),
in Appendices \ref{a:generictriplet} and \ref{a:genericgap}
we work out 
respectively the Bethe-Salpeter and gap equations for 
semiconducting tubes in the presence of the magnetic field.

\section{Effective-mass approximation}\label{s:model}

In this section we recall the $\bm{k\cdot p}$ 
theory of electronic $\pi$-states in single-wall 
carbon nanotubes (CNTs) according to 
Ando.\cite{Ajiki1993,Ando1997,Ando2006}

Carbon nanotubes may be thought of as wrapped sheets of graphene,
hence nanotube electronic states are built from those of graphene
after imposing suitable boundary conditions. Here 
we focus on single-particle levels lying close to 
K (isospin $\tau=1$) or K$'$ ($\tau=-1$) points
in graphene's reciprocal space, 
where Dirac cones' apexes touch. These two apexes are
the Fermi surface of undoped graphene. 
The envelope functions $\bm{F}^{\tau}(\bm{r})$ of CNT single-particle states
are two-component spinors, each component being the 
wave function amplitude on one of the two sublattices. These envelopes
obey the $\bm{k\cdot p}$ equations of graphene, 
\begin{equation}
\gamma(\sigma_x \hat{k}_x + \tau \sigma_y \hat{k}_y)\bm{F}^{\tau}(\bm{r})
= \varepsilon \bm{F}^{\tau}(\bm{r}),
\end{equation}
plus the additional boundary condition along the tube circumference:
\begin{equation}
\bm{F}^{\tau}(\bm{r+L}) = \bm{F}^{\tau}(\bm{r})
\exp \left[ 2\pi i (\varphi - \tau\frac{\nu}{3})\right].
\end{equation}
Here $\bm{L}$ is the chiral vector in the circumference direction of the
CNT, $\nu = 0,\pm 1$ is the chirality index that depends on the microscopic
structure of the CNT ($\nu = 0$ for metals and $\nu = \pm 1$ 
for semiconductors),
$\varphi = \phi / \phi_0$ is the ratio of the magnetic 
flux $\phi$ through the tube cross section 
to the magnetic flux quantum $\phi_0 = ch/e$, 
$\gamma$ is graphene's band parameter, 
$\sigma_x$ and $\sigma_y$ are the Pauli matrices,
$\varepsilon$ is the single-particle energy,
$\hat{k}_x = -i\partial /\partial x$ is the wave vector operator
along the circumference direction $x$ and 
$\hat{k}_y = -i\partial /\partial y$ acts on the tube axis coordinate $y$.

\begin{figure}
\setlength{\unitlength}{1 cm}
\begin{picture}(8.5,11.0)
\put(0.0,0.0){\epsfig{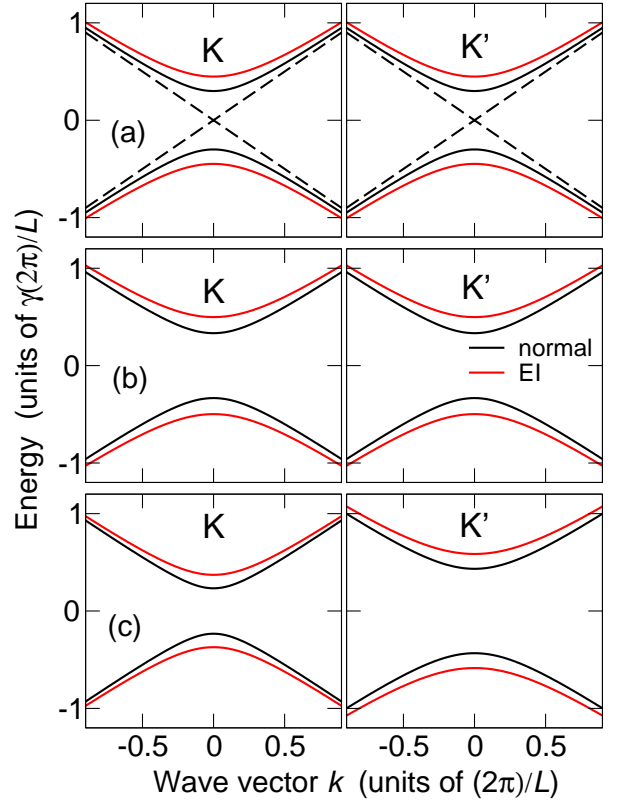}}
\end{picture}
\caption{(color online)
Dispersion of electronic energy levels 
of single-wall carbon nanotubes
close to the charge neutrality point. The dependence of the
energy on wave vector $k$
in both valleys K (left column) and K$'$ (right column) is
shown. Black (red [gray]) lines represent noninteracting
(EI quasiparticle) energy bands $\varepsilon^{\tau}(k)$
[$E^{\tau}(k)$].
(a) Metallic tube in the presence of the magnetic 
flux $\varphi=0.3$. 
Here $\varphi = \phi / \phi_0$ is the ratio of the magnetic 
flux $\phi$ through the tube cross section 
to the magnetic flux quantum $\phi_0 = ch/e$. 
The dashed line is the dispersion with $\varphi=0$.
(b) Semiconducting tube with $\nu=1$ and $\varphi=0$.
(c) Semiconducting tube with $\nu=1$ and $\varphi=0.1$.
As the field along the tube axis is increased the gap in valley K$'$ increases
while the gap in valley K decreases.
Quasiparticle energies are obtained  
from their weak-coupling
expressions for the sake of comparison among all panels,
with $a(2\pi/L) = 1$ and 
$w_2 = 100\gamma (2\pi/L)$.
\label{gap_binding}}
\end{figure}

The energy bands are specified by the index
$\alpha = (n, \ell)$ plus
the wave vector $k$ in the axis direction, where $n$ is an integer
and $\ell = c$, $v$ denotes either the conduction ($\ell = c$) 
or the valence band ($\ell = v$).
The wave functions in the K valley are
\begin{equation}
\bm{F}^{\text{K}}_{\alpha k}  (\bm{r}) 
=
\bm{\xi}^{\text{K}}_{\alpha k}(x)  
\frac{1}{\sqrt{A}} \exp{(iky)} ,
\end{equation}
where $A$ is the CNT length and the wave function 
$\bm{\xi}^{\text{K}}_{\alpha k}(x)$ for the motion along
the circumference direction is  
\begin{equation}
\bm{\xi}^{\text{K}}_{\alpha k}  (x) =
\frac{1}{\sqrt{L}} \exp{[ik_{\nu}(n)  x]} 
\bm{F}^{\nu}_{\alpha k},  
\label{eq:spinor} 
\end{equation}
with $L=\left|\bm{L}\right|$ being the tube circumference.
In Eq.~\eqref{eq:spinor} the transverse wave vector 
$k_{\nu}(n)$ depends on the magnetic flux,
\begin{equation}
k_{\nu}(n) =
\frac{2\pi}{L} \left( n + \varphi -
\frac{\nu}{3} \right),
\label{eq:k_x} 
\end{equation}
and the spinor 
$\bm{F}^{\nu}_{\alpha k}$ is  
a unit vector with a $\bm{k}$-dependent
phase between the two sublattice components,
\begin{equation}
\bm{F}^{\nu}_{\alpha k}  =
\frac{1}{\sqrt{2}} 
{  b_{\nu}(n,k) \choose  s_{\alpha}  } ,
\label{eq:F} 
\end{equation}
where
\begin{equation}
b_{\nu}(n,k) =
\frac{ k_{\nu}(n) -ik  }{\sqrt{ k^2_{\nu}(n) + k^2     }} ,
\label{eq:b} 
\end{equation}
and
$s_{\alpha}=\pm 1$ for conduction and valence bands, respectively.
The corresponding energy is
\begin{equation}
\varepsilon^{\text{K}}_{\alpha}(k) = s_{\alpha}  \gamma
\sqrt{ k^2_{\nu}(n) + k^2     } ,
\label{eq:energy} 
\end{equation}
which is reckoned from the charge neutrality point (Fig.~\ref{gap_binding}). 

The wave function in the K$'$ valley is
\begin{equation}
\bm{F}^{\text{K}'}_{\alpha k}  (\bm{r}) 
=
\bm{\xi}^{\text{K}'}_{\alpha k}(x)  
\frac{1}{\sqrt{A}} \exp{(iky)} ,
\end{equation}
with
\begin{equation}
\bm{\xi}^{\text{K}'}_{\alpha k}  (x) =
\frac{1}{\sqrt{L}} \exp{[ik_{-\nu}(n)  x]} 
\bm{F}^{-\nu *}_{\alpha k},  
\label{eq:spinorKpri} 
\end{equation}
whereas its energy is
\begin{equation}
\varepsilon^{\text{K}'}_{\alpha}(k) = s_{\alpha}  
\gamma \sqrt{ k^2_{-\nu}(n) + k^2     } .
\label{eq:energyKpri} 
\end{equation}

\section{Bethe-Salpeter equation for spinless excitons}\label{s:allband}

In this section we recall Ando's results for spinless excitons 
in a single valley\cite{Ando1997} to validate a simpler
two-band model that is at the basis of our further development.

The exciton wave function $\left|u\right>$ of
zero center-of-mass momentum in the---say---K valley is
\begin{equation}
\left| u \right> = \sum_n \sum_k \psi_n(k)
\hat{c}^{\text{K}+}_{n,c,k}
\hat{c}^{\text{K}}_{n,v,k} \left|g\right>,
\label{eq:excitonWF} 
\end{equation}
where $\left|g\right>$ is the ground state of the intrinsic CNT
with all $v$ bands filled and $c$ bands empty,
$\psi_n(k)$ is the $n$th component of
the exciton wave function in the reciprocal space,
and the operator $\hat{c}^{\text{K}+}_{n,c,k}$ creates 
an electron in valley K having conduction-band index $n$ 
and momentum $k$. There are no off-diagonal contributions
with different band indices in \eqref{eq:excitonWF} 
since these are forbidden by the symmetry of
Coulomb interaction [cf.~\eqref{eq:BSEintravalley}].

The Bethe-Salpeter equation for the exciton wave function
$\psi_n(k)$ of eigenvalue $\varepsilon_u$ is: 
\begin{eqnarray}
&& \varepsilon_u \psi_n(k)  = 
\left[
\varepsilon^{\text{K}}_{n,c}(k) -\varepsilon^{\text{K}}_{n,v}(k) 
+ 
\Delta\varepsilon^{\text{K}}_n(k)
\right] \psi_n(k) 
\nonumber \\
&& -  \sum_{m,q} 
V_{ (n,c,k;m,c,k+q)(m,v,k+q;n,v,k) }
\,\psi_m(k+q) .
\label{eq:BSEintravalley}
\end{eqnarray}
Here $ \varepsilon^{\text{K}}_{n,c}(k)   
-\varepsilon^{\text{K}}_{n,v}(k) = 
2\gamma \left[ k^2_{\nu}(n) + k^2\right]^{1/2}$ is the energy cost for
creating a noninteracting electron of momentum $k$ 
in the $n$th conduction band and a hole the $n$th valence band, 
$\Delta\varepsilon^{\text{K}}_n(k)$ is the sum of electron
and hole quasiparticle self-energies due to interaction,
and $V_{ (n,c,k;m,c,k+q)(m,v,k+q;n,v,k) }$ 
is the screened Coulomb matrix element that scatters
different electron-hole pairs,
binding the electron and the hole. The above quantities
were evaluated by Ando within the
random phase approximation and the eigenvalue problem
\eqref{eq:BSEintravalley} was solved numerically, as detailed 
in Ref.~\onlinecite{Ando1997}. 

\begin{figure}
\setlength{\unitlength}{1 cm}
\begin{picture}(8.5,5.7)
\put(0.5,0.0){\epsfig{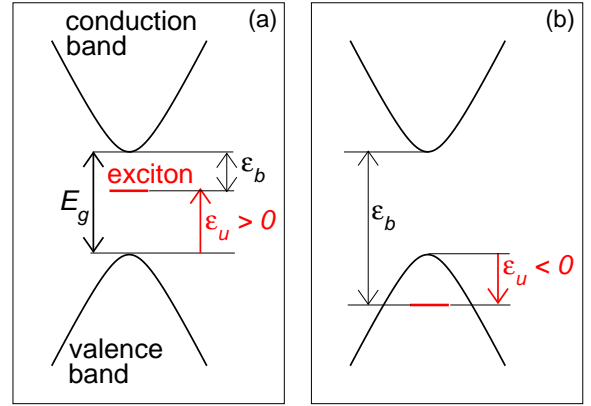}}
\end{picture}
\caption{(color online)
Excitonic instability.  
(a) In the ordinary case the excitation energy $\varepsilon_u$  
of the exciton is positive and smaller than the quasiparticle gap
$E_g$. The exciton binding energy $\varepsilon_b$ is smaller than $E_g$.
(b) An excitonic instability occurs for $\varepsilon_u<0$ ($\varepsilon_b
>E_g$), 
which drives the transition to the excitonic insulator phase.
\label{inst}}
\end{figure}

Ando considered generic CNTs exhibiting a gap, i.e.,
both semiconductors ($\nu=\pm 1$)
and metals ($\nu=0$) in the presence of 
the magnetic field ($\varphi\neq 0$). 
He found that the terms related to Coulomb interaction appearing 
on the right hand side of the Bethe-Salpeter equation
\eqref{eq:BSEintravalley}---the self-energy
$\Delta\varepsilon^{\text{K}}$ and
the matrix element $V$---are large taken separately, because of
the reduced dimensionality of the CNT and its poor screening.
However, these two terms cancel out almost exactly, hence
the binding energy lowers the exciton level below the quasiparticle band edge 
of continuum states while the self-energy lifts the band edge above its
noninteracting value.
The net effect is that the energy of the lowest exciton level,
$\varepsilon_u$, is slightly blueshifted with respect to the bare 
energy gap, $2\gamma\left|k_{\nu}(0)\right|$, and redshifted with 
respect to the interacting energy gap, $E_g$.
Since the exciton  energy is always positive, $\varepsilon_u>0$, 
Ando concluded that an excitonic instability never occurs [Fig.~\ref{inst}(a)]. 
A similar result was inferred by Hartmann and coworkers
on the basis of a semianalytical model.\cite{Hartmann2011} 

To derive a simpler exciton model, we use the key result by Ando 
that the lowest exciton energy $\varepsilon_u$ is homogeneous and
almost linear with the gap $E_g$,
as shown in Fig.~5 of Ref.~\onlinecite{Ando1997} 
for metallic CNTs as a function
of the magnetic field that opens the gap.
The rationale is that the smaller the band gap the stronger the screening,
which makes the exciton binding energy approximately proportional to the gap. 
Besides, $\varepsilon_u$ weakly depends on the strength of Coulomb interaction, 
$(e^2/\epsilon_r L)/(2\pi \gamma/L)$,
as seen in Fig.~4 of Ref.~\onlinecite{Ando1997} showing that 
$\varepsilon_u$ is almost constant for reasonable values of 
$(e^2/\epsilon_r L)/(2\pi \gamma/L) > 0.1$.
Here, the quantity
$(e^2/\epsilon_r L)/(2\pi \gamma/L)$ is dimensionless, with
$(e^2/\epsilon_r L)/(2\pi \gamma/L) = 0.3545 / \epsilon_r$,
$\epsilon_r$ being the unknown static dielectric constant describing 
contributions from states far from the charge 
neutrality point, 
and $2\pi \gamma/L$ being the energy unit.

On the basis of the results plotted
in Figs.~4 and 5 of Ref.~\onlinecite{Ando1997} that we have recalled above, 
we approximate $\varepsilon_u$ as
\begin{equation}
\varepsilon_u \approx \beta\, E_g,
\label{eq:my}
\end{equation}
where $E_g$ is the band gap renormalized by Coulomb interaction---possibly
in the presence of the magnetic field---and
$\beta$ is a fraction of the unity, $\beta \approx 0.8$.

\section{Two-band model for spinless excitons}\label{s:twoband}

Next we introduce a simpler two-band model that reproduces 
the numerical result \eqref{eq:my} for the lowest exciton 
state.
This model will be the starting point of our theory.

\begin{figure}
\setlength{\unitlength}{1 cm}
\begin{picture}(8.5,5.7)
\put(0.5,0.0){\epsfig{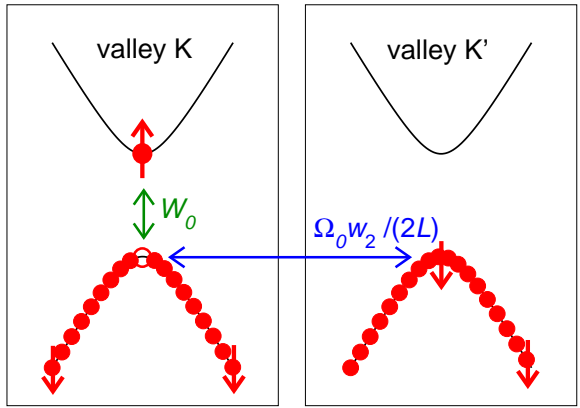}}
\end{picture}
\caption{(color online)
Two-band model for the
lowest triplet exciton. $W_0$ is the intravalley interaction
energy and $\Omega_0w_2/2L$ is the intervalley exchange interaction.
The arrows represent electron spins. Here the exciton spin projection
is $S_z=1$.
\label{f:triplet}}
\end{figure}

We consider only the lowest conduction and highest valence band,
labeled by $n=0$ for moderate values of $\varphi$, so the exciton 
wave function is
\begin{equation}
\left| u \right> = \sum_k \psi(k)
\hat{c}^{\text{K}+}_{c,k}
\hat{c}^{\text{K}}_{v,k} \left|g\right>,
\label{eq:exciton_my} 
\end{equation}
with the shorthands $\psi(k)\equiv \psi_{n=0}(k)$
and 
$\hat{c}^{\text{K}+}_{c,k}\equiv \hat{c}^{\text{K}+}_{0,c,k}$.
The Bethe-Salpeter equation takes the form
\begin{eqnarray}
&&\left[
2\gamma\sqrt{k_{\nu}(0)^2 + k^2   }
+ 
\Delta\varepsilon(k)
\right] \psi(k) 
 -  \sum_{q} 
V_{ k;k+q }
\,\psi(k+q) \nonumber \\
&&
= \quad \varepsilon_u \psi(k). 
\label{eq:BSE_2b_intra}
\end{eqnarray}
Furthermore, we assume that: (i) the effect of the self-energy  
$\Delta\varepsilon(k)$ may be included into the band structure by
renormalizing the band parameter $\gamma$ 
(ii) the screened Coulomb matrix element 
$V_{k;k+q}$ does not depend on momenta,
$V_{k;k+q}\equiv W_0/A$ with $W_0>0$ (cf.~Fig.~\ref{f:triplet}).
The former assumption implies that the value of $\gamma$
should be estimated through a quasiparticle calculation, 
like the GW technique.
The latter approximation
provides a contact attractive interaction 
that captures the essential physical features
of excitons in CNTs.\cite{Spataru2004,Maultzsch2005} 

We proceed Fourier transforming 
\eqref{eq:BSE_2b_intra} in real space,
\begin{equation}
2\gamma\sqrt{k_{\nu}(0)^2 - \frac{d^2}{dy^2}   }\;
\psi(y) 
 -  W_0\,\delta(y) \,\psi(y)
= \varepsilon_u \psi(y), 
\label{eq:BSE_2b_real}
\end{equation}
with $\psi(y)$ being the exciton wave function for the relative
motion of the electron-hole pair.
The square root operator appearing in \eqref{eq:BSE_2b_real}
is a symbolic expression for the power series
\begin{eqnarray}
&&\sqrt{k_{\nu}(0)^2 - \frac{d^2}{dy^2}   } =
\left|k_{\nu}(0)\right|\nonumber\\ && + \quad 
\left|k_{\nu}(0) \right|\sum_{n=1}^{\infty}{1/2 \choose n}    
\frac{(-1)^n}{\left| k_{\nu}(0)\right|^{2n} }   \frac{d^{(2n)}}{dy^{(2n)}}.   
\label{eq:sqrt}
\end{eqnarray}
It is tempting to keep only the second derivative in the expansion
on the right hand side of
\eqref{eq:sqrt}, thus recovering the familiar Wannier equation for a massive 
exciton in an usual semiconductor.\cite{Kozlov1965} However,
later we will be interested in exploring
the excitonic instability
(Fig.~\ref{inst}), which is the critical regime that one reaches
at the frontier of the domain of convergence 
of the power series \eqref{eq:sqrt}. Therefore we 
fully take into account the relativistic dispersion of the exciton---a
peculiar feature of CNTs---and solve exactly 
\eqref{eq:BSE_2b_real}. 
Our approach allows for a solution also 
in the supercritical regime $\varepsilon_u < 0$,
as we will discuss in Sec.~\ref{s:instability},
contrary to the method illustrated in
Ref.~\onlinecite{Hartmann2011}. 
Both relativistic and nonrelativistic treatments
provide the same results for small ratioes of $W_0$ to $\gamma$.

As in the nonrelativistic case, the solution of \eqref{eq:BSE_2b_real} 
is a bound state whose
wave function $\psi(y)$ takes the form
\begin{equation}
\psi(y) = \sqrt{\kappa}\, \exp( -\kappa \left|y\right| ),
\label{eq:BSEsolution}
\end{equation}
and whose energy $\varepsilon_u$ is smaller than the gap
$E_g = 2\gamma\left|k_{\nu}(0)\right|$,
\begin{equation}
\varepsilon_u = 2\gamma\sqrt{k_{\nu}(0)^2 - \kappa^2 } .
\label{eq:sqrt_e}
\end{equation}
Here the exciton inverse decay length
$\kappa>0$ has to be determined through
the proper boundary condition at the origin. This condition
is obtained by integrating both sides
of \eqref{eq:BSE_2b_real} over an infinitesimal 
interval containing the origin, providing
\begin{equation}
-4\gamma 
\left|k_{\nu}(0)\right| \sum_{n=1}^{\infty}{1/2 \choose n}    
\frac{(-1)^n}{\left| k_{\nu}(0)\right|^{2n} }   
\kappa^{2n-1}   = W_0,
\label{eq:boundary}
\end{equation}
which may be resummed as
\begin{equation}
\gamma\sqrt{k_{\nu}(0)^2 - \kappa^2 } = \gamma\left|k_{\nu}(0)\right| 
-\frac{\kappa W_0}{4}.
\label{eq:boundary2}
\end{equation}
Resolving \eqref{eq:boundary2} for $\kappa$ one obtains 
\begin{equation}
\kappa = 
\frac{8\gamma \left|k_{\nu}(0)\right|  W_0}{16\gamma^2 + W_0^2},
\label{eq:kappa}
\end{equation}
therefore the exciton energy is:
\begin{equation}
\varepsilon_u = 2\gamma\left|k_{\nu}(0)\right|  
\frac{16\gamma^2 - W_0^2}{16\gamma^2 + W_0^2}.
\label{eq:energy_onevalley}
\end{equation}
As expected, the attractive contact
interaction $W_0$ lowers $\varepsilon_u$ below the band gap $E_g$.

The value of $W_0$ is determined imposing the constraint
\eqref{eq:my},
which provides
\begin{equation}
W_0 = 
4\,\sqrt{ \frac{1 - \beta}{1 + \beta} }\;\gamma \approx 1.33 \, \gamma   .
\label{eq:W0estimate}
\end{equation}
The latter estimate corresponds to $\beta = 0.8$.
The parameter $\beta$  
weakly depends on the strength of Coulomb interaction
for realistic values of $\epsilon_r$.
We see that $\gamma$ is the only energy scale that appears in the equation
of motion \eqref{eq:BSE_2b_real} for a Dirac exciton in a single valley.

\section{Exchange interaction}\label{s:intervalley} 

Whereas in Secs.~\ref{s:allband}
and \ref{s:twoband} we have focused on excitons whose 
electron and hole constituents occupy the K valley (KK exciton),
we may consider as well excitons in the 
K$'$ valley 
(K$'$K$'$ exciton) and excitons made of the electron in the K valley and
hole in the K$'$ valley (KK$'$ exciton) or vice versa (K$'$K exciton).
Including the spin degree of freedom, there is a total amount of 
sixteen different excitons made of an electron in the valley $\tau$
with spin $\sigma$ and a hole
in the valley $\tau'$ with spin $\sigma'$, 
that we label as $\left|(\tau,\sigma)(\tau',\sigma')\right>$,
with $\tau$, $\tau'$ = K or K$'$ and $\sigma$, $\sigma'$ = $\uparrow$ or
$\downarrow$ (here we comply with the notation by Ando\cite{Ando2006}
that the spin of the
hole is denoted by that of the missing electron in the valence band).

These excitons, which are all
degenerate in the absence of the magnetic field, may be further classified
in terms of the total spin as singlet and triplet excitons.
There are four singlet excitons, 
\begin{equation}
^1\!\left|\tau\tau'\right> = 
\frac{1}{\sqrt{2}}\left[\left|(\tau,\uparrow)(\tau',\uparrow)\right> +
\left|(\tau,\downarrow)(\tau',\downarrow)\right>\right],
\end{equation}
and four triplet excitons that are separately threefold degenerate due
to the different spin projections $S_z = +1, 0, -1$,
\begin{eqnarray}
&& ^3\!\left|\tau\tau',+1\right>  = 
\left|(\tau,\uparrow)(\tau',\downarrow)\right>,\nonumber\\
&& ^3\!\left|\tau\tau',0\right>  =  
\frac{1}{\sqrt{2}}\left[\left|(\tau,\uparrow)(\tau',\uparrow)\right> -
\left|(\tau,\downarrow)(\tau',\downarrow)\right>\right], \nonumber\\
&& ^3\!\left|\tau\tau',-1\right>  = 
\left|(\tau,\downarrow)(\tau',\uparrow)\right>.
\end{eqnarray}
In the following we take
into account the orbital coupling
with the magnetic field through the transverse wave vector $k_{\nu}(0)$
but neglect the small Zeeman coupling  
lifting spin degeneracies $S_z$. 

Ando has showed that the sixteen-fold degeneracy of
the exciton manifold is lifted 
by the small short-range part of Coulomb interaction, which is not
included in the $\bm{k\cdot p}$ 
theory\cite{Ando2006,Secchi2013} and
is responsible for spin and valley exchange.
Intervalley exchange interaction
splits singlet states into the bonding and antibonding of 
$^1\!\left|\text{KK}\right>$ and $^1\!\left|\text{K}'\text{K}'\right>$   
and two degenerate $^1\!\left|\text{K}\text{K}'\right>$ and  
$^1\!\left|\text{K}'\text{K}\right>$.  
Triplet excitons split in the same way
although each triplet is three-fold degenerate.
The overall ordering and energy splitting is determined by two
exchange parameters, $w_1$ and
$w_2$. 

The generic lowest exciton state turns out to be 
the triplet `bonding' exciton,
\begin{equation}
\frac{1}{\sqrt{2}}\left[^3\!\left|\text{KK},S_z\right> +
^3\!\left|\text{K}'\text{K}',S_z\right> \right] \quad 
\text{with}\quad S_z=-1,0,1,
\label{eq:dark}
\end{equation}
which is optically inactive (Fig.~\ref{f:triplet}). 
The energy of this bonding exciton
is lower then that of the single-valley KK or K$'$K$'$ excitons 
due to intervalley exchange. This prediction agrees with
available state-of-art fully ab-initio 
calculations.\cite{Spataru2004,Maultzsch2005}
Since we are concerned with
ground-state properties only, 
we will limit our study to the triplet exciton  
\eqref{eq:dark} 
focusing on its specific equation of motion.

The short-range part of Coulomb interaction 
has both an intravalley and an intervalley
contribution, respectively $\hat{V}^{(1)}$ and $\hat{V}^{(2)}$. 
The matrix element of $\hat{V}^{(1)}$  has the form
\begin{equation}
\Omega_0\,w_1
\!\int\! d\bm{r}\! \left[
\bm{F}^{\tau}_{\alpha_1 k_1}\!(\bm{r})^{\dagger} \sigma_z\,
\bm{F}^{\tau}_{\alpha_2 k_2}(\bm{r})\right] \left[
\bm{F}^{\tau'}_{\alpha_3 k_3}\!(\bm{r})^{\dagger} \sigma_z\,
\bm{F}^{\tau'}_{\alpha_4 k_4}(\bm{r})\right] ,
\label{eq:w_1int}
\end{equation}
where $\Omega_0 = (\sqrt{3}/2)a^2$ is the area of graphene unit cell,
$a=2.46$ {\AA} is the lattice constant, and $w_1>0$ characterizes 
intravalley interaction strength.
The matrix element of $\hat{V}^{(2)}$  is
\begin{eqnarray}
&& \Omega_0\,w_2\!\int\! d\bm{r}\! \Big[
\bm{F}^{\tau A}_{\alpha_1 k_1}\!(\bm{r})^{*} 
\bm{F}^{\tau' A}_{\alpha_2 k_2}(\bm{r}) 
\bm{F}^{\tau' A}_{\alpha_3 k_3}\!(\bm{r})^{*} 
\bm{F}^{\tau A}_{\alpha_4 k_4}(\bm{r}) \nonumber \\
&& + \qquad 
\bm{F}^{\tau B}_{\alpha_1 k_1}\!(\bm{r})^{*} 
\bm{F}^{\tau' B}_{\alpha_2 k_2}(\bm{r}) 
\bm{F}^{\tau' B}_{\alpha_3 k_3}\!(\bm{r})^{*} 
\bm{F}^{\tau B}_{\alpha_4 k_4}(\bm{r}) 
\Big] ,
\label{eq:w_2int}
\end{eqnarray}
where $\tau \neq \tau'$, the apex $A$ ($B$) labels the first (second)
spinorial sublattice component, and $w_2>0$ characterizes 
intervalley interaction strength (cf.~Fig.~\ref{f:triplet}). 
We expect that the effect of screening
on matrix elements \eqref{eq:w_1int} and \eqref{eq:w_2int} is smaller 
than that on the conventional long-range
Coulomb terms discussed in Sec.~\ref{s:allband},
hence we neglect it.\cite{Ando2006}

The exchange terms $\hat{V}^{(1)}$ and $\hat{V}^{(2)}$
pertinent to  
the two-band model introduced in Sec.~\ref{s:twoband} 
take the form
\begin{eqnarray}
\hat{V}^{(1)} & = & 
\frac{\Omega_0\,w_1}{2AL}\sum_{\tau\tau'}
\sum_{\alpha\beta\alpha'\beta'}\sum_{kk'q}\sum_{\sigma\sigma'}\nonumber\\
&\times &
V^{(1)}_{(\tau,\alpha,k+q;\tau\beta,k)(\tau',\beta',k';\tau',\alpha',k'+q)} 
\nonumber \\
&\times &
\hat{c}^{\tau +}_{\alpha,k+q,\sigma}
\hat{c}^{\tau' +}_{\beta',k',\sigma'}
\hat{c}^{\tau'}_{\alpha',k'+q,\sigma'}
\hat{c}^{\tau}_{\beta,k,\sigma},
\label{eq:w_1int_2} 
\end{eqnarray}
and
\begin{eqnarray}
\hat{V}^{(2)} & = & 
\frac{\Omega_0\,w_2}{2AL}
\sum_{\tau\neq\tau'}
\sum_{\alpha\beta\alpha'\beta'}\sum_{kk'q}\sum_{\sigma\sigma'}\nonumber\\
&\times &
V^{(2)}_{(\tau',\alpha,k+q;\tau\beta,k)(\tau,\beta',k';\tau',\alpha',k'+q)} 
\nonumber \\
&\times &
\hat{c}^{\tau' +}_{\alpha,k+q,\sigma}
\hat{c}^{\tau +}_{\beta',k',\sigma'}
\hat{c}^{\tau'}_{\alpha',k'+q,\sigma'}
\hat{c}^{\tau}_{\beta,k,\sigma},
\label{eq:w_2int_2} 
\end{eqnarray}
where 
$\hat{c}^{\tau}_{\beta,k,\sigma}$
destroys an electron in the valley $\tau$ with momentum $k$ and spin $\sigma$
occupying either the conduction  
($\beta=c$) or valence ($\beta=v$) band $n=0$. 
Explicitly,
the matrix elements are
\begin{eqnarray}
&&V^{(1)}_{(\tau,\alpha,k+q;\tau\beta,k)(\tau',\beta',k';\tau',\alpha',k'+q)}
 \nonumber \\
&&= \quad \frac{1}{4}
\left[b^*_{\nu} (0,k+q) b_{\nu}(0,k)-s_{\alpha}s_{\beta}\right]\nonumber\\
&& \times \quad 
\left[b_{-\nu} (0,k') b^*_{-\nu}(0,k'+q)-s_{\beta'}s_{\alpha'}\right]
\label{eq:w_1int_3} 
\end{eqnarray}
and
\begin{eqnarray}
&&V^{(2)}_{(\tau',\alpha,k+q;\tau\beta,k)(\tau,\beta',k';\tau',\alpha',k'+q)} 
 \nonumber \\
&&= \quad \frac{1}{4}
\Big[ b_{-\nu}(0,k+q) b_{\nu}(0,k) b^*_{\nu}(0,k') b^*_{-\nu}(0,k'+q) 
\nonumber\\
&& + \quad 
s_{\alpha}s_{\beta}    s_{\beta'}s_{\alpha'} 
\Big].
\label{eq:w_2int_3} 
\end{eqnarray}
These matrix elements
are further simplified taking $k$, $k'$, $q\approx 0$, in the spirit 
of the $\bm{k\cdot p}$ method. Then \eqref{eq:w_1int_3} is 1 if 
$(\alpha \neq \beta)\wedge 
(\alpha' \neq \beta')$ and zero otherwise, \eqref{eq:w_2int_3} 
is 1/2 if $s_{\alpha}s_{\beta}    s_{\beta'}s_{\alpha'}=1$
and zero otherwise. 
The term
\eqref{eq:w_2int_2}, responsible for intervalley
exchange, is the only short-range contribution relevant to the dynamics of the 
triplet exciton \eqref{eq:dark}, as we discuss below.

\section{Bethe-Salpeter equation for the lowest triplet 
exciton}\label{s:BSEtriplet}

In this section we extend the two-band model of Sec.~\ref{s:twoband} 
adding spin and valley degrees of freedom 
to treat the lowest triplet exciton \eqref{eq:dark}.
The generic
wave function of this exciton is 
\begin{equation}
\left| u \right> = \sum_{\tau,k} \psi_{\tau} (k)
\hat{c}^{\tau +}_{c,k,\uparrow}
\hat{c}^{\tau}_{v,k,\downarrow} \left|g\right>.
\label{eq:exciton_dark} 
\end{equation}
Here we consider both 
metallic and semiconducting nanotubes  
and choose the spin projection
$S_z = 1$ for the sake of clarity, as illustrated in Fig.~\ref{f:triplet}.

The short-range terms that enter
the equation of motion must scatter the electron-hole pairs
$\hat{c}^{\tau +}_{c,k,\uparrow}
\hat{c}^{\tau}_{v,k,\downarrow} \left|g\right>$ that span
the triplet exciton subspace. 
The intravalley operator $\hat{V}^{(1)}$
unaffects this subspace---at least at the lowest order---except
for a small negative constant term lowering the exciton energy that  
may be neglected.\cite{Ando2006} On the other hand, the intervalley
operator $\hat{V}^{(2)}$ transfers electron-hole pairs from one 
valley to the other one, whereas the diagonal term is the same
for both ground and exciton states (Fig.~\ref{f:triplet}). 
The resulting equations of motion, 
Fourier transformed in real space, are:
\begin{eqnarray}
\varepsilon_u \psi_{\text{K}} (y) 
&=& 2\gamma\sqrt{k_{\nu}(0)^2 - \frac{d^2}{dy^2}   }\;
\psi_{\text{K}} (y) 
 -  W_0\,\delta(y) \,\psi_{\text{K}} (y)
\nonumber \\
&&-\quad \frac{\Omega_0\,w_2}{2L} \delta(y) \,\psi_{\text{K}'} (y) \nonumber \\
\varepsilon_u \psi_{\text{K}'} (y) 
&=& 2\gamma\sqrt{k_{-\nu}(0)^2 - \frac{d^2}{dy^2}   }\;
\psi_{\text{K}'} (y) 
 -  W_0\,\delta(y) \,\psi_{\text{K}'} (y)
\nonumber \\
&& -\quad \frac{\Omega_0\,w_2}{2L} \delta(y) \,\psi_{\text{K}} (y). 
\label{eq:BSE_triplet}
\end{eqnarray}

Equations \eqref{eq:BSE_triplet}
provide the bound state of a two-component
massive exciton with a relativistic dispersion.
In the relative-motion frame of \eqref{eq:BSE_triplet} 
two types of attractive scattering potentials occur at the origin---the
coordinate at which 
the electron and the hole share the same position along the CNT axis. 
One contact term is due to intravalley Coulomb interaction 
with strength
$W_0$ and the other one to intervalley exchange with strength
$\Omega_0 w_2 / 2 L$, as shown in Fig.~\ref{f:triplet}.

\begin{figure}
\setlength{\unitlength}{1 cm}
\begin{picture}(8.5,5.7)
\put(0.5,0.0){\epsfig{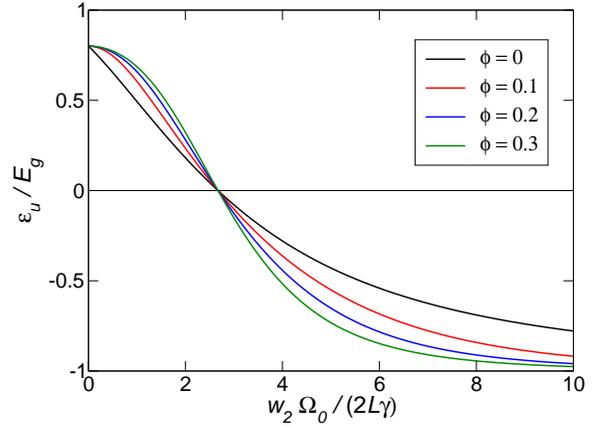}}
\end{picture}
\caption{(color online)
Energy $\varepsilon_u$ of the lowest triplet exciton vs 
exchange interaction strength $(\Omega_0 w_2 / 2 L)/\gamma$
for different magnetic field fluxes $\varphi$. 
The black line represents the energy of either a metallic tube ($\nu=0$)
at any field or a semiconducting tube ($\nu=\pm1$)
at zero field. Colored [gray] lines correspond to $\nu=\pm1$ and finite field.
Here $\varphi = \phi / \phi_0$ is the ratio of the magnetic 
flux $\phi$ through the tube cross section 
to the magnetic flux quantum $\phi_0 = ch/e$. 
The energy $\varepsilon_u$ is in units of the quasiparticle gap $E_g$,
with $E_g=2\gamma\left|k_{\nu=0}(0)\right|$ for a metal
and $E_g=\min( 2\gamma\left|k_{\nu=1}(0)\right|, 
2\gamma\left|k_{\nu=-1}(0)\right| )$ for a semiconductor.
At the critical value of interaction $\Omega_0 w_2 / 2 L+W_0=4\gamma$
the exciton energy becomes negative (here $W_0 = 1.33 \gamma$).
\label{supercritical}}
\end{figure}

The solution of \eqref{eq:BSE_triplet} is straightforward
in the case the gaps in the two valleys are identical, 
$\left|k_{\nu}(0)\right| = \left|k_{-\nu}(0)\right|$
(black line in Fig.~\ref{supercritical}).
This occurs for either a metallic CNT ($\nu = 0$) at any field
[Fig.~\ref{gap_binding}(a)]
or a semiconducting tube ($\nu = \pm 1$) at zero field
[Fig.~\ref{gap_binding}(b)]
(see Appendix \ref{a:generictriplet} for the generic solution). 
Then the components of the exciton 
in the two valleys are identical as well, 
$ \psi_{\text{K}} (y) = \psi_{\text{K}'} (y)$, their explicit form 
being given by
\eqref{eq:BSEsolution} after replacing $W_0$ with 
$W_0 +  \Omega_0 w_2 / 2 L$. The exciton energy is
\begin{equation}
\varepsilon_u = 2\gamma\left|k_{\nu}(0)\right|  
\frac{16\gamma^2 - \left[W_0
+\left(\Omega_0 w_2 / 2 L\right) 
\right]^2}
{16\gamma^2 + \left[W_0 +\left(\Omega_0 w_2 / 2 L
\right) \right]^2},
\label{eq:energy_twovalley}
\end{equation}
which is the same as the spinless result
\eqref{eq:energy_onevalley}
except for the key difference that the exchange interaction strength
$\Omega_0 w_2 / 2 L$ now adds to $W_0$.

For semiconductors at finite field the exciton energy is
obtained numerically, as shown in Fig.~\ref{supercritical} 
(colored [gray] lines).
As the mismatch between the band edges of valleys K and K$'$ increases 
with the field ($\varphi$ increases) the exciton binding energy decreases
($\varepsilon_u$ increases). 

\section{Excitonic instability}\label{s:instability}

Whereas the intravalley interaction strength $W_0$ 
weakly depends on tube parameters,
the exchange interaction strength
$\Omega_0 w_2 / 2 L$ 
that enters the exciton energy
\eqref{eq:energy_twovalley}
depends on the inverse of tube circumference $L$
as well as on the microscopic details
of the interaction
potential and exciton wave function through the
parameter $w_2$ (Ref.~\onlinecite{Ando2006}). 
As shown in Fig.~\ref{supercritical}, at the critical value
\begin{equation}
W_0 + \frac{\Omega_0 w_2}{2L} = 4\gamma
\label{eq:critical}
\end{equation}
the excitation energy $\varepsilon_u$ 
goes to zero,
thus the `normal' ground state $\left|g\right>$ becomes unstable 
against the spontaneous formation of triplet excitons.

This scenario, proposed in the Sixties by Mott,\cite{Mott1961} 
Keldysh,\cite{Keldysh1964}  
and Kohn\cite{Kohn1967} 
among others, is illustrated in Fig.~\ref{inst}(b). 
In the supercritical regime $\varepsilon_u<0$
the ground state rearranges itself into
the long-sought  
`excitonic insulator' (EI) phase.
The EI phase, which is pictorially depicted in Fig.~\ref{f:clock}(b),
is a permanent
condensate of excitons at thermodynamic equilibrium
exhibiting insulating behavior.
The quasiparticle gap (red [gray] lines in Figs.~\ref{gap_binding}
and \ref{f:clock})
is widened with respect to the normal ground state due to the emergence
of a many-body contribution $\Delta$ reminescent of the exciton 
binding energy [see Eq.~\eqref{eq:Eqp} below].  

Using the estimate by Ando\cite{Ando2006} 
of $w_2\approx$ 4 eV, 
Eq.~\eqref{eq:W0estimate},
and taking $\gamma = 5.39$ eV$\cdot$\AA\ 
we obtain as a critical value of the tube circumference
$L = 0.73$ \AA, which is at least one order of magnitude smaller than realistic 
values. However, quoting Ando,\cite{Ando2006} 
`It is worth being pointed out that
the parametrization into $w_1$ and $w_2$ is much more general although
their actual values can be different from those estimated above
[$\ldots$] we should leave $w_1$ and $w_2$ rather
as adjustable parameters to be determined experimentally'.
In much the same spirit, we propose that the observable properties 
of the excitonic insulator should be used to extract the value of $w_2$.

Beyond the critical value of exchange interaction, 
$W_0 + \Omega_0 w_2/2L > 4\gamma$, 
the exciton energy $\varepsilon_u$ becomes negative
according to \eqref{eq:energy_twovalley}, which is at odds with
the square-root dependence 
of the kinetic energy in
\eqref{eq:BSE_2b_real}
only allowing for
positive values of $\varepsilon_u$
[cf.~Eq.~\eqref{eq:sqrt_e}].
This issue is related to the problem of supercritical fields in
quantum electrodynamics,\cite{Greiner1985} which has received
a great deal of attention in the context of 
graphene\cite{Pereira2007,Shytov2007a,Shytov2007b,Wang2013} and carbon 
nanotubes.\cite{Hartmann2011,Maksym2013}  
In the present context, we note that the sum of the infinite
series 
defining the kinetic energy operator 
on the rhs of \eqref{eq:sqrt}
is a double valued function.
Taking the negative square root and repeating
the calculation of Sec.~\ref{s:twoband} 
we obtain that the formulae 
\eqref{eq:kappa}, \eqref{eq:energy_onevalley}, \eqref{eq:energy_twovalley}
for both the exciton inverse decay length $\kappa$
and energy $\varepsilon_u$ are analitically continued in the
supercritical regime ($\varepsilon_u < 0$).

The exciton energy $\varepsilon_u$ 
in the whole range of exchange interaction is
plotted in Fig.~\ref{supercritical}. The 
binding energy 
$\varepsilon_b = \left|2\gamma \left|k_{\nu}
(0)\right|-\varepsilon_u\right|$ 
in the supercritical regime 
becomes larger than the gap $2\gamma\left|k_{\nu}(0)\right|$, 
reaching its maximum allowed value 
of twice the gap, $4\gamma\left|k_{\nu}(0)\right|$, at infinite exchange
interaction ($\varepsilon_u/E_g=-1$).
Therefore, contrary to the case of relativistic electron states 
in superheavy atoms, the exciton bound level never
merges the antiparticle continuum 
lying at the bottom of the forbidden energy gap, 
located at  
$-2\gamma\left|k_{\nu}(0)\right|$    
(not to be confused with the top of the valence band).
This prediction is in striking contrast with the conclusions
of Hartmann and coworkers.\cite{Hartmann2011}

\section{Excitonic insulator}\label{s:ei}

In this section we build up the Hartree-Fock theory of the 
EI phase of carbon nanotubes, which significantly departs  
from the treatment of usual semiconductors\cite{Kozlov1965,Kohn1967,Halperin1967,Kohn1968,Volkov1975}
due to the relativistic character and chirality of electrons.
We include in our two-band Hamiltonian
$\hat{H}$ only those terms responsible for the excitonic instability,
\begin{equation}
\hat{H} = \hat{H}_0 + \hat{V}_{\text{intra}} + \hat{V}_{\text{inter}},
\end{equation}
where $\hat{H}_0$ is the noninteracting term,
\begin{equation}
\hat{H}_0 = \sum_{\tau,\alpha, k,\sigma}
\varepsilon^{\tau}_{\alpha}\!(k)
\,\hat{c}^{\tau +}_{\alpha,k,\sigma}
\hat{c}^{\tau }_{\alpha,k,\sigma},
\end{equation}
$\hat{V}_{\text{intra}}$ is the intravalley interaction term introduced
in Sec.~\ref{s:twoband}, 
\begin{equation}
\hat{V}_{\text{intra}} 
= \frac{W_0}{A}
\sum_{\tau}\sum_{kk'q}\sum_{\sigma\sigma'}
\hat{c}^{\tau +}_{c,k+q,\sigma}
\hat{c}^{\tau +}_{v,k'-q,\sigma'}
\hat{c}^{\tau }_{v,k',\sigma'}
\hat{c}^{\tau }_{c,k,\sigma},
\end{equation}
and $\hat{V}_{\text{inter}}$ is the intervalley exchange term discussed
in Sec.\ref{s:intervalley},
\begin{equation}
\hat{V}_{\text{inter}} 
= \frac{\Omega_0 w_2}{2AL}
\sum_{\tau\neq \tau'}\sum_{kq}\sum_{\sigma\sigma'}
\hat{c}^{\tau' +}_{c,k+q,\sigma}
\hat{c}^{\tau +}_{v,k,\sigma'}
\hat{c}^{\tau' }_{v,k+q,\sigma'}
\hat{c}^{\tau }_{c,k,\sigma}.
\end{equation}

Furthermore, we introduce the intraband one-particle Green function,
\begin{equation}
G^{\alpha}_{\sigma\sigma'}(\tau k, t)=-i\hbar^{-1}\left<T\left\{
\hat{c}^{\tau }_{\alpha,k,\sigma}(t)
\hat{c}^{\tau +}_{\alpha,k,\sigma'}(0)\right\}\right>,
\label{eq:G}
\end{equation}
as well as the `anomalous' interband Green function,
\begin{equation}
F^{\alpha\beta}_{\sigma\sigma'}(\tau k, t)=-i\hbar^{-1}
\left<T\left\{
\hat{c}^{\tau }_{\alpha,k,\sigma}(t)
\hat{c}^{\tau +}_{\beta,k,\sigma'}(0)\right\}\right>.
\label{eq:FGreen}
\end{equation}
The latter is zero for the noninteracting ground state but takes a finite value
in the EI phase,
pointing to electron-hole interband correlations 
(here $\alpha\beta= cv$ or $vc$).
In \eqref{eq:G} and \eqref{eq:FGreen} 
$T$ is the time-ordering operator and $\left<\ldots\right>$ the
quantum average over the ground state.\cite{Abrikosov1975}

Whereas
the intraband Green function
is diagonal in the spin space as the EI ground state 
has no net spin magnetization, 
\begin{equation}
G^{\alpha}_{\sigma\sigma'}(\tau k, t)= 
\delta_{\sigma\sigma'}G^{\alpha}(\tau k, t),
\end{equation}
the interband Green function is spin-polarized along the (arbitrary)
direction $\bm{n}$, 
\begin{equation}
F^{\alpha\beta}_{\sigma\sigma'} (\tau k, t)
=\left(
\bm{n\cdot \sigma}\right)_{\sigma\sigma'} 
F^{\alpha\beta} (\tau k, t),
\label{eq:Fsigma}
\end{equation}
with $\bm{\sigma}$ being the vector formed by the three Pauli matrices
and $\bm{n}$ a constant unit vector.
This may be understood as 
$F^{cv}_{\sigma\sigma'} (\tau k, 0+)$ is proportional to the
wave function in reciprocal space of the condensate of triplet excitons 
whose spins are polarized along $\bm{n}$. This condensate has no macroscopic magnetization
but exhibits a periodic modulation of the spin density within each unit cell of
the honeycomb lattice (antiferromagnetic spin density wave).\cite{Kozlov1965,Halperin1967} 
Whereas the long range order of the spin density wave 
is destroyed by quantum and thermal fluctuations 
in an indefinitely long tube, here we assume the size of the sample 
is comparable to the spin-spin correlation length. 

We obtain the Fourier-transformed
quantities $G^{\alpha}(\tau k, \omega)$ 
and 
$F^{\alpha\beta} (\tau k, \omega)$ 
from their Heisenberg equations of motion
after applying the Hartree-Fock decoupling scheme.\cite{Kohn1967}
The system of equations involving $G^{c}(\tau k, \omega)$ 
is
\begin{eqnarray}
\left[ \hbar\omega - \varepsilon^{\tau}_{c}\!(k)\right]
G^{c}(\tau k, \omega) - \Delta(\tau k)\,
F^{vc} (\tau k, \omega)  &=& 1 \nonumber \\
\left[ \hbar\omega - \varepsilon^{\tau}_{v}\!(k)\right]
F^{vc} (\tau k, \omega) 
- \Delta^*\!(\tau k) \,
G^{c}(\tau k, \omega) &=& 0,
\label{eq:FGsystem}
\end{eqnarray}
where we have introduced the gap function $\Delta(\tau k)$
defined through the equation
\begin{eqnarray}
\Delta^*\!(\tau k) &=& \frac{i\hbar}{A} \sum_q
\Big[ W_0 \,F^{vc} (\tau\, k+q, t=0+) \nonumber \\
&&+\quad\frac{\Omega_0 w_2}{2L} F^{vc} (-\tau\, k+q, t=0+) 
\Big],
\label{eq:Delta}
\end{eqnarray}
with $-\tau$ labeling the valley different from $\tau$.
The equations involving $G^{v}(\tau k, \omega)$ do not provide
additional information.
Solving system
\eqref{eq:FGsystem} for $G^c$ and $F^{vc}$ in terms of $\Delta$,
we obtain
\begin{equation}
G^{c}(\tau k, \omega) = \frac{ \hbar\omega - \varepsilon^{\tau}_{v}\!(k) 
}{ \left[ \hbar\omega - \varepsilon^{\tau}_{c}\!(k)\right] 
\left[ \hbar\omega - \varepsilon^{\tau}_{v}\!(k)\right] -\left| 
\Delta(\tau k) \right|^2               } ,
\label{eq:Gsolved}
\end{equation}
\begin{equation}
F^{vc}(\tau k, \omega) = \frac{  \Delta^*\!(\tau k) 
}{ \left[ \hbar\omega - \varepsilon^{\tau}_{c}\!(k)\right] 
\left[ \hbar\omega - \varepsilon^{\tau}_{v}\!(k)\right] -
\left| \Delta(\tau k) \right|^2               
} .
\label{eq:Fsolved}
\end{equation}

The poles of Green functions \eqref{eq:Gsolved} and 
\eqref{eq:Fsolved} are the (spin-degenerate) quasiparticle energies 
$E^{\tau}\!(k)$ (Ref.~\onlinecite{Abrikosov1975}). These bands   
are plotted in Figs.~\ref{gap_binding} and
\ref{f:clock} (red [gray] lines), 
their explicit form being given by
\begin{eqnarray}
E^{\text{K} }(k) & = & \pm  
\sqrt{ \gamma^2 k^2_{\nu}(0) 
+ \gamma^2 k^2  
+ \left| \Delta(\text{K}  k) \right|^2                  } , \nonumber \\
E^{\text{K}'}\!(k) & = & \pm  
\sqrt{ \gamma^2 k^2_{-\nu}(0) + \gamma^2 k^2  
+ \left| \Delta(\text{K}' k) \right|^2                  } ,
\label{eq:Eqp} 
\end{eqnarray}
where the sign plus and minus refers to electrons and holes, respectively.
The result \eqref{eq:Eqp} shows that 
the valley-dependent gap function
$\left| \Delta(\tau k) 
\right|$ 
adds quadratically to the noninteracting half-gap, 
$\gamma \left| k_{\pm\nu}(0) \right|$. Substituting 
\eqref{eq:Fsolved} into 
\eqref{eq:Delta} and integrating over the frequency we
obtain the gap equation,
\begin{eqnarray}
\Delta(\tau k) 
& = & 
\frac{1}{A} 
\sum_q
\Big[ W_0 \frac{\Delta(\tau\, k+q ) }{2\left|E^{\tau}\!(k+q)\right| } 
\nonumber \\
&&+\quad\frac{\Omega_0 w_2}{2L}  
\frac{\Delta(-\tau\, k+q ) }{2\left|E^{-\tau}\!(k+q)\right| } 
\Big],
\label{eq:gap}
\end{eqnarray}
which implicitly provides $\left| \Delta(\tau k) \right|$. 
This equation 
has the same structure as the gap equation of BCS
theory of superconductivity.\cite{BCS1957,deGennes1999}
Excluding the trivial 
noninteracting solution $\Delta(\tau k)=0$,
a finite value of $\Delta(\tau k)$
points to the stability of the EI phase with respect to the
noninteracting ground state. The phase of $\Delta(\tau k)$ 
is arbitrary and the energy is independent of it.

\section{Solution of the gap equation}\label{s:gap}

\begin{figure}
\setlength{\unitlength}{1 cm}
\begin{picture}(8.5,6.5)
\put(0.3,0.0){\epsfig{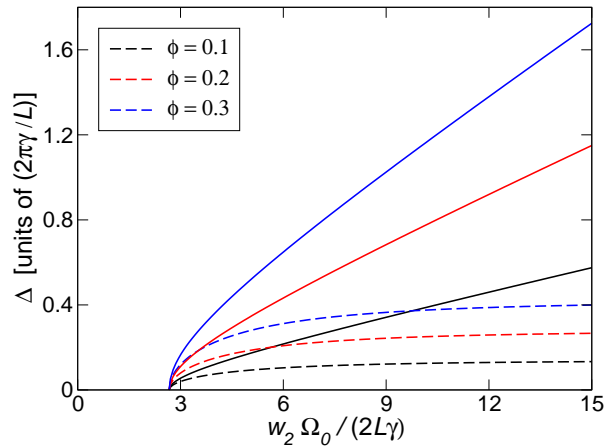}}
\end{picture}
\caption{(color online)
Excitonic gap $\Delta$ of a nominally metallic tube ($\nu=0$)
vs interaction strength $(\Omega_0 w_2 / 2L) /\gamma$
at zero temperature for different values of the magnetic flux $\varphi$.
The dashed lines are the values obtained by a weak-coupling expansion.
Here $\varphi = \phi / \phi_0$ is the ratio of the magnetic 
flux $\phi$ through the tube cross section 
to the magnetic flux quantum $\phi_0 = ch/e$ and
$W_0 = 1.33\gamma$.
\label{Delta_phase}}
\end{figure}

To solve the gap equation \eqref{eq:gap}
we assume that the excitonic
gap function $\Delta(\tau k)$ 
is independent from $k$, $\Delta(\tau k)
\equiv \Delta(\tau)$,  
and rewrite \eqref{eq:gap} as
\begin{eqnarray}
&& 2\sqrt{ \gamma^2 k^2_{\nu}(0) 
+ \gamma^2 k^2  
+ \left| \Delta(\text{K}  ) \right|^2  }   
\varphi^{\text{K}}(k) \nonumber \\
&& =  
\frac{1}{A} 
\sum_q
\left[ W_0\, \varphi^{\text{K}}(k+q)  
+\frac{\Omega_0 w_2}{2L}  \varphi^{\text{K}'}\!(k+q)  
\right], \nonumber \\
&& 2\sqrt{ \gamma^2 k^2_{-\nu}(0) 
+ \gamma^2 k^2  
+ \left| \Delta(\text{K}'  ) \right|^2  }   
\varphi^{\text{K}'}(k) \nonumber \\
&& =  
\frac{1}{A} 
\sum_q
\left[ W_0\, \varphi^{\text{K}'}\!(k+q)  
+\frac{\Omega_0 w_2}{2L}  \varphi^{\text{K}}(k+q)  
\right],
\label{eq:gap2}
\end{eqnarray}
with the position
$\varphi^{\tau}(k) = \Delta(\tau)/2\left|E^{\tau}\!(k)\right| $. 
We see that for a vanishing value of the excitonic gap---at the border 
of the EI phase, $\Delta(\tau) = 0+$,
the gap equation coincides with the Bethe-Salpeter equation 
\eqref{eq:BSE_triplet} in momentum space for the triplet exciton 
of zero energy. This confirms that at the critical interaction strength
\eqref{eq:critical}
the nanotube undergoes a transition 
to the EI phase (Fig.~\ref{Delta_phase}).

\subsection{Weak coupling}

There are two ways of solving the gap equation \eqref{eq:gap2}.
The first method is valid only at weak coupling, i.e.,
the excitonic gap is much smaller than the noninteracting gap,
$\left|\Delta(\tau)\right|\ll \gamma\left|k_{\nu}(0)\right|$,
hence we may expand
the square root entering \eqref{eq:gap2} 
in terms of $\left|\Delta(\tau)\right|$ (Ref.~\onlinecite{Kozlov1965}).   
In the case of either
metallic tubes or semiconducting tubes at zero field 
the gaps in the two valleys are
equal, $\Delta(\text{K}) = \Delta(\text{K}') \equiv \Delta$, 
hence at the lowest order
\eqref{eq:gap2} reduces to   
\begin{eqnarray}
&& 2\gamma \sqrt{ k^2_{\nu}(0) 
+ k^2  }   
\varphi(k) 
- \frac{1}{A} 
\sum_q
\left[ W_0  
+\frac{\Omega_0 w_2}{2L}\right]  \varphi(k+q) \nonumber \\ 
&&=\quad 
-\frac{\left|\Delta\right|^2}{\gamma \left|k_{\nu}(0)\right| } \varphi(k),
\label{eq:gap3}
\end{eqnarray}
with $\varphi^{\tau}\equiv \varphi$. 
This is identical to the Bethe-Salpeter 
equation \eqref{eq:BSE_triplet} for a triplet 
exciton of energy $-\left|\Delta\right|^2/\gamma \left|k_{\nu}(0)\right|$.
Equating this latter energy to \eqref{eq:energy_twovalley} and
solving for $\left|\Delta\right|$, we obtain
\begin{equation}
\left|\Delta\right| = \sqrt{2}\gamma\left|k_{\nu}(0)\right|  
\sqrt{
\frac{\left[W_0
+\left(\Omega_0 w_2 / 2 L\right) 
\right]^2-16\gamma^2}
{ \left[W_0 +\left(\Omega_0 w_2 / 2 L
\right) \right]^2 + 16\gamma^2  
}
}
\label{eq:DeltaEI}
\end{equation}
for $W_0 +\Omega_0 w_2 / 2 L \ge 4\gamma$ 
and $\left|\Delta\right|=0$ otherwise (dashed lines in Fig.~\ref{Delta_phase}).
The generic weak-coupling case of a valley-dependent excitonic gap 
$\Delta(\tau)$ for semiconducting tubes
at finite field is worked out in Appendix \ref{a:genericgap}
and illustrated in Fig.~\ref{Delta_K_Kpr_phase}.

\begin{figure}
\setlength{\unitlength}{1 cm}
\begin{picture}(8.5,6.5)
\put(0.3,0.0){\epsfig{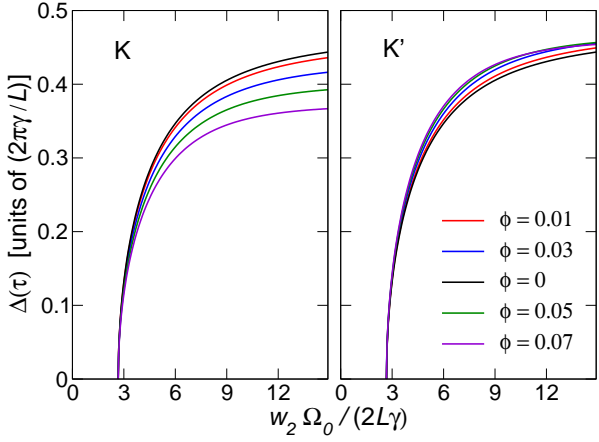}}
\end{picture}
\caption{(color online)
Excitonic gap $\Delta(\tau)$ of a semiconducting tube ($\nu=1$)
vs interaction strength $(\Omega_0 w_2 / 2L) /\gamma$
for different values of the magnetic flux $\varphi$.
The data refer to zero temperature and weak coupling.
Left panel: valley K. Right panel: valley K$'$.
Here $\varphi = \phi / \phi_0$ is the ratio of the magnetic 
flux $\phi$ through the tube cross section 
to the magnetic flux quantum $\phi_0 = ch/e$ and
$W_0 = 1.33\gamma$.
\label{Delta_K_Kpr_phase}}
\end{figure}

The dependence of $\left|\Delta\right|$ on both the exchange interaction 
strength $\Omega_0 w_2 / 2 L$ 
and noninteracting gap width 
$2\gamma\left|k_{\nu}(0)\right|$ tuned by the field
is displayed in Fig.~\ref{Delta_phase} (dashed lines).  
The excitonic gap $\left|\Delta\right|$ is zero below the
critical threshold of intervalley exchange interaction and increases up to
the maximum value of $\sqrt{2}\gamma\left|k_{\nu}(0)\right|$ at 
strong interaction strength. 
Beyond the critical threshold, the EI phase is stable for any size of 
the noninteracting bandgap. The latter is tuned 
by the magnetic field, the excitonic gap vanishing with the 
noninteracting gap.  

This scaling of excitonic and noninteracting 
gaps---a consequence of the relativistic
dispersion of Dirac electrons---is in striking contrast with the behavior
of the EI gap predicted for ordinary semiconductors.\cite{Kozlov1965}    
In this latter case the onset of EI phase occurs when,
decreasing the size of noninteracting 
gap e.g.~by applying pressure to the solid, 
the band gap equals the exciton binding energy, which only depends on the
exciton mass and dielectric constant. 

\subsection{Strong coupling}

In the strong-coupling limit of arbitrary exchange interaction strength,
$\Omega_0 w_2 / 2 L$, 
we directly perform the sum over $q$ in \eqref{eq:gap},
limiting ourselves to the case of valley-independent $\Delta$.
This sum exhibits an unphysical ultraviolet divergence 
that originates from the
assumed independence of $\Delta$ from $k$.
Therefore, we introduce the cutoff $k_c$ in the summation \eqref{eq:gap}.  
The fact that $\Delta $ vanishes at the critical interaction strength
\eqref{eq:critical} 
provides a constrain that fixes the value of $k_c$.
The result is
\begin{equation}
k_c = \sinh{\frac{\pi}{2}}
\left|k_{\nu}(0)\right|
\approx 2.301\left|k_{\nu}(0)\right| ,
\label{eq:cutoff}
\end{equation}
scaling with the noninteracting gap.
Note that
\eqref{eq:cutoff} compares with the cutoff value $2\pi/L$ taken in the 
numerical calculations of Ref.~\onlinecite{Ando1997}.
After performing the integration of \eqref{eq:gap2},
the generic form of $\Delta$ is
\begin{equation}
\Delta = \gamma \left|k_{\nu}(0)\right|
\left[ 
\frac{
\sinh{\left(\frac{\pi}{2}\right)}^2 }
{ \sinh{\left(\frac{2\pi\gamma}
{W_0 + \Omega_0 w_2 /(2L)     }
\right)}^2
} - 1
\right]^{1/2}    
\label{eq:Delta_strong}
\end{equation}
for $W_0 + \Omega_0 w_2 /(2L)>4\gamma$ and $\Delta = 0$ otherwise.

The strong-coupling value of $\Delta$ given by \eqref{eq:Delta_strong}
is compared with the weak-coupling prediction 
in Fig.~\ref{Delta_phase} (solid versus dashed lines,
respectively). Whereas the two families of curves overlap close
to the critical interaction strength, the strong-coupling value
approximately doubles its weak-coupling counterpart 
around $w_2\Omega_0/(2L\gamma)\sim 6$, increasing unbounded 
with interaction strength.

\begin{figure}
\setlength{\unitlength}{1 cm}
\begin{picture}(8.5,6.5)
\put(0.3,0.0){\epsfig{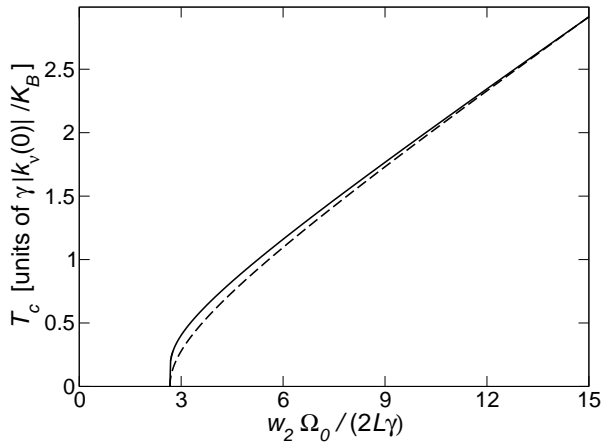}}
\end{picture}
\caption{
Critical temperature $T_c$ 
vs interaction strength $(\Omega_0 w_2 / 2L) /\gamma$ (solid line).
The temperature scales linearly with $\gamma\left|k_{\nu}(0)\right|$, i.e.,
half the noninteracting gap. 
The noninteracting tube is either
metallic ($\nu=0$) in the presence of the magnetic field or
semiconducting ($\nu=\pm 1$) in the absence of the field.
Here $K_B$ is Boltzmann constant
and $W_0 = 1.33\gamma$. For comparison, the excitonic gap 
at $T=0$, $\Delta(T=0)/(\gamma\left|k_{\nu}(0)\right|K_B)$,
rescaled by the factor 0.5067,
is shown as a dashed line.  
\label{T_c}}
\end{figure}

The effect of temperature $T$ on the excitonic gap 
$\Delta$ may be evaluated straightforwardly, paralleling the
procedure of BCS theory.\cite{deGennes1999} The $T$-dependent
gap equation is
\begin{eqnarray}
&&\frac{2\pi\gamma}{W_0 + \Omega_0 w_2 /(2L) }
=\!\!\!
\int_0^{\sinh{(\pi/2)} }
\frac{ dt }{\left( 1 + \tilde{\Delta}^2 + t^2   \right)^{1/2}    } 
\nonumber\\
&\qquad&\times \qquad \tanh{\left[ \frac{\tilde{\beta} }{2}  
\left( 1 + \tilde{\Delta}^2 + t^2  \right)^{1/2} \right]  } ,
\label{eq:DeltaT}
\end{eqnarray}
where we have defined the reduced quantities 
$\tilde{\Delta}   = \Delta(T)/(\gamma\left|k_{\nu}(0)\right|)$   
and
$\tilde{\beta}  = \beta\, \gamma\left|k_{\nu}(0)\right|$,
with $\beta = 1 /( K_B T)$ and $K_B$ is the Boltzmann constant. 
The critical temperature $T_c$ is obtained 
putting $\tilde{\Delta}=0$ into
\eqref{eq:DeltaT} 
and then solving numerically
the corresponding equation, which provides the relation between
$T_c$ and interaction strength. 

The outcome 
is shown in Fig.~\ref{T_c}. As it is evident from the
comparison between $T_c$ (solid line)
and the excitonic gap at zero temperature $\Delta(T=0)$, rescaled
by the factor 0.5067 (dashed line),
both curves share approximately
the same dependence on $\Omega_0 w_2 / 2 L$, at least for strong interaction. 
Therefore, the equation 
\begin{equation}
K_BT_c \approx 0.507\; \Delta(T=0)
\end{equation}
provides a useful estimate of the effect
of temperature.
Since typical small-gap semiconducting tubes exhibit transport
gaps---possibly of excitonic origin---of the order of tens of meV 
whereas measurements are performed 
around $T \approx 100$ mK,\cite{Deshpande2009,Steele2013}
one may neglect the effect of temperature at the first instance.

\begin{figure}
\setlength{\unitlength}{1 cm}
\begin{picture}(8.5,6.5)
\put(0.3,0.0){\epsfig{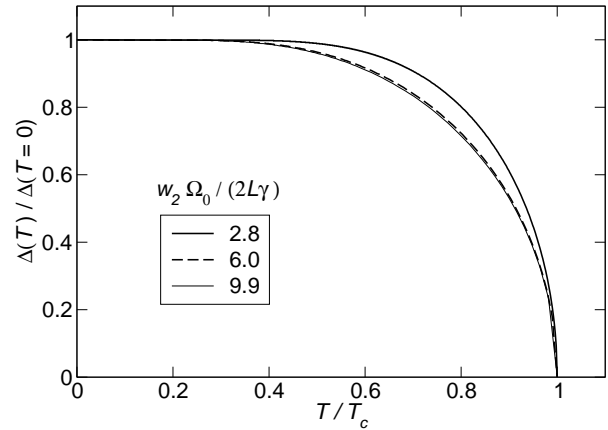}}
\end{picture}
\caption{
Normalized excitonic gap $\Delta(T)/\Delta(T=0)$
vs normalized temperature $T/T_c$ for different values of
interaction strength $(\Omega_0 w_2 / 2L) /\gamma$.
The noninteracting tube is either
metallic ($\nu=0$) in the presence of the magnetic field or
semiconducting ($\nu=\pm 1$) in the absence of the field.
\label{DeltaTvsT}}
\end{figure}

This is confirmed by the temperature dependence of $\Delta(T)$,
which is illustrated in Fig.~\ref{DeltaTvsT}. Similarly to the
gap in BCS theory, close to $T=0$ the excitonic gap
$\Delta(T)$ exhibits a large
plateau whereas in the neighborhood of 
$T_c$ it drops with a square-root dependence.  
Therefore, $\Delta(T)$ is substantially unaffected by $T$ in a broad
range of cryogenic temperatures relevant for experiments.
In the remaining part of the paper we will neglect   
temperature effects.

\section{Excitonic enhancement of the quasiparticle 
magnetic moment}\label{s:anomalous}

From the knowledge of the quasiparticle energy $E^{\tau}\!(k)$ 
it is straightforward
to compute the magnetic moment per quasiparticle, $\mu$,  
defined as the negative slope of the
quasiparticle energy as a function
of the magnetic field,\cite{Ashcroft1976}
\begin{equation}
\mu = -\left(\frac{\partial E^{\tau}\!(k)}{\partial B}\right)_{B=0},
\label{eq:mudef}
\end{equation}
to be understood as the left- or right-hand limit  
in the presence of a cusp.
The magnetic moment may be measured through single-electron
tunneling spectroscopy.\cite{Minot2004}

\begin{figure}
\setlength{\unitlength}{1 cm}
\begin{picture}(8.5,6.5)
\put(0.3,0.0){\epsfig{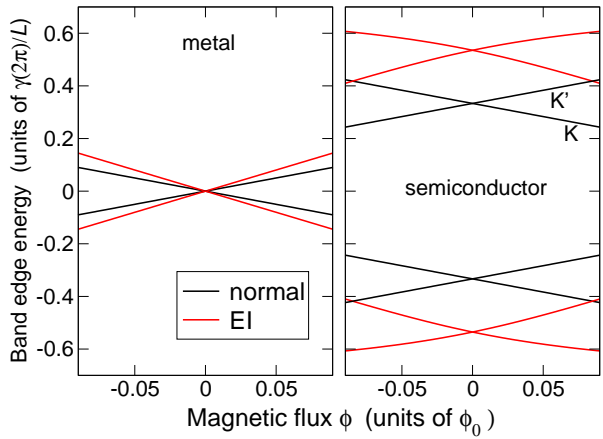}}
\end{picture}
\caption{(color online)
Dependence of EI quasiparticle (noninteracting)
band edges $E^{\tau}\!(k=0)$ [$\varepsilon^{\tau}\!(k=0)$]
on the magnetic flux $\varphi$.
Left panel: metal ($\nu=0$). Right panel: semiconductor ($\nu=1$).
Here $\varphi = \phi / \phi_0$ is the ratio of the magnetic 
flux $\phi$ through the tube cross section 
to the magnetic flux quantum $\phi_0 = ch/e$, 
$a(2\pi/L) = 1$, and $w_2 = 150\gamma (2\pi/L)$.
For the sake of comparison the weak-coupling results are used in
both panels.
\label{edge_phi}}
\end{figure}

The dependence of quasiparticle band edges $E^{\tau}\!(k=0)$ on $\varphi$
is illustrated in Fig.~\ref{edge_phi} 
for both metallic (left panel) and semiconducting (right panel) tubes. 
The magnitudes of the slopes of quasiparticle bands (red [gray] lines)
are enhanced with respect to noninteracting bands (black lines). 
While EI band edges of both metallic and semiconducting tubes have comparable 
slopes at the origin, they exhibit differences at large values 
of $\varphi$ as the band edges of semiconducting tubes deviate from linearity.

The case of
metallic nanotubes ($\nu=0$) allows for an analytical treatment as
the excitonic gap $\Delta$ is the same in both valleys, 
$E(k)\equiv E^{\tau}\!(k)$. 
Recalling that 
the magnetic field $B$ along the tube axis enters both equations 
\eqref{eq:Eqp} and
\eqref{eq:Delta_strong}
through the field-dependent transverse wave vector [Eq.~\eqref{eq:k_x}],
\begin{equation}
k_{\nu=0}(0) =
\frac{BLe}{2 c h} ,
\label{eq:k_B} 
\end{equation}
after derivation we obtain
\begin{equation}
\mu = \pm \mu_0
\left[
\frac{
\sinh{\left(\frac{\pi}{2}\right)} } 
{ \sinh{\left(\frac{2\pi\gamma}
{W_0 + \Omega_0 w_2 /(2L)     }
\right)}
}
\right]
.
\label{eq:mu} 
\end{equation}

Here the sign plus (minus) corresponds to the (anti)parallel 
orientation of the magnetic moment of the quasiparticle with 
respect to the magnetic field, 
and $\mu_0 = e v_F R /2c$ is the 
semiclassical value
of the magnetic moment of the electron rotating around the tube 
circumference with  
Fermi velocity $v_F = \gamma / \hbar$ (Ref.~\onlinecite{Minot2004}).

\begin{figure}
\setlength{\unitlength}{1 cm}
\begin{picture}(8.5,6.5)
\put(0.3,0.0){\epsfig{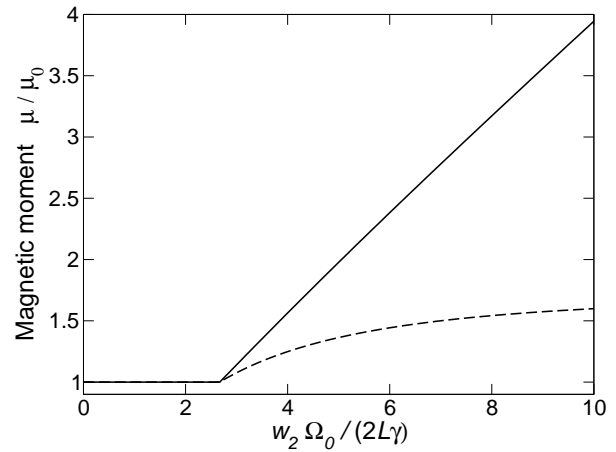}}
\end{picture}
\caption{Quasiparticle magnetic moment $\left|\mu\right|/\mu_0$
of a nominally metallic tube
vs exchange interaction strength $(\Omega_0 w_2 / 2 L)/\gamma$
at zero temperature. The dashed curve is the weak-coupling prediction.
Here $W_0 = 1.33 \gamma$ and $\mu_0$ is the semiclassical
estimate of the magnetic moment.
\label{f:mu}
}
\end{figure}

In the noninteracting phase the term enclosed by square brackets
in \eqref{eq:mu} 
is the unity, hence one recovers the semiclassical result
$\mu = \pm \mu_0$. In the EI phase this value
is increased by the factor in parentheses, 
in a fashion proportional to the excitonic gap $\Delta$---at least for
strong interaction.
The dependence of $\mu$ on intervalley exchange interaction 
strength $(\Omega_0 w_2 / 2 L)/\gamma$ in the EI phase
is illustrated in Fig.~\ref{f:mu}, which highlights the excitonic enhancement
with respect to $\mu_0$. The peculiar scaling of $\mu$ 
with $R$ is the experimental hallmark of exciton
condensation, as further discussed in Sec.~\ref{s:exp} 
(cf.~Fig.~\ref{f:magmom}).
Note that the weak-coupling estimate
of $\mu$ (dashed line in Fig.~\ref{f:mu}) tends to the horizontal 
asymptote $\mu = \pm \sqrt{3}\mu_0$ 
whereas the strong-coupling prediction (solid line) increases 
indefinitely with intervalley exchange interaction.  

\section{Weak paramagnetism of the exciton 
condensate}\label{s:weak}

To shed light onto the unusual value of the magnetic moment 
of quasiparticles we consider the magnetization of the EI ground state,
focusing on nominally metallic tubes ($\nu=0$) for the sake of clarity.
We compute only the 
orbital magnetization $M$ along the nanotube axis since the spin
contribution is negligible,
the Zeeman spin term coupling with the field being much smaller than the
orbital coupling term.\cite{Minot2004}

The total magnetization $M$ is the sum of the magnetic
moments of filled one-electron levels, each moment being weighted 
by the level occupancy. 
All levels share approximately the same absolute value of the magnetic moment, 
$\mu_0$, whose sign is opposite for conduction and valence bands,
respectively. 
This may be understood in a semiclassical picture,\cite{Minot2004} 
as the transverse wave vector 
$k_{\nu=0}(0)\propto B$ along the
circumferential direction $x$ is the same for conduction and 
valence states whereas the corresponding group velocities 
$1/\hbar\left(\partial \varepsilon_{\alpha} / \partial k\right)_x$ 
have alternate signs.
Therefore, all electrons in the conduction band
rotate---say---anticlockwise around the tube circumference whereas those
in the valence band rotate clockwise, as shown pictorially
in Fig.~\ref{f:clock}(a). 
At finite wave vector $k$
the magnetic moment $\mu_0(k)$ slightly departs from $\mu_0$
due to the change in the group velocity:
\begin{equation}
\mu_0(k)=\frac{\mu_0}{\sqrt{1 + k^2/k_{\nu=0}^2(0) } }.
\label{eq:group}
\end{equation}

In terms of intraband Green functions, the magnetization is
\begin{eqnarray}
M & = & -i\hbar \sum_{\tau k} \mu_0(k) \sum_{\sigma}  \Big[
G^v_{\sigma\sigma} (\tau k, t = 0-) \nonumber\\
& \quad & -G^c_{\sigma\sigma} (\tau k, t = 0-)
\Big].
\label{eq:Mdef} 
\end{eqnarray}
Integrating over the frequency, we obtain
\begin{equation}
M = 4 \sum_k \mu_0(k) \left(u_k^2 - v_k^2\right),
\label{eq:M}
\end{equation}
where the coherence factors $u_k$ and $v_k$ are defined
in analogy with the BCS theory of superconductivity:
\begin{eqnarray}
u_k^2  & = &
\frac{1}{2}\left(1 + \frac{\sqrt{\gamma^2k^2 + \gamma^2k_{\nu=0}^2(0)}}
{\sqrt{\gamma^2k^2 + \gamma^2k_{\nu=0}^2(0) + \left|\Delta\right|^2}}\right),
\nonumber \\
v_k^2  & = &  1 - u_k^2.
\label{eq:uv_def}
\end{eqnarray}

The quantities $u_k^2$ and $v_k^2$ are the populations
of valence- and conduction-band levels, respectively.
In the noninteracting ground state the excitonic
gap $\Delta$ vanishes, 
hence $u_k^2=1$, $v_k^2=0$, 
that is the valence band is filled and the conduction band empty,
so one obtains $M_0=\mu_0 ALB (e/hc)$ (the subscript 
0 identifies the noninteracting phase).
Therefore, the tube is a paramagnet having susceptibility per unit length
$\mu_0 L(e/hc)$ 
and magnetization 
proportional to the semiclassical dipole $\mu_0$ times
the number of magnetic flux quanta piercing the tube surface $AL$. 
This scenario is illustrated in Fig.~\ref{f:clock}(a),
with all electrons carring their magnetic dipole 
aligned with the field $B$.

\begin{figure}
\setlength{\unitlength}{1 cm}
\begin{picture}(8.5,6.5)
\put(0.3,0.0){\epsfig{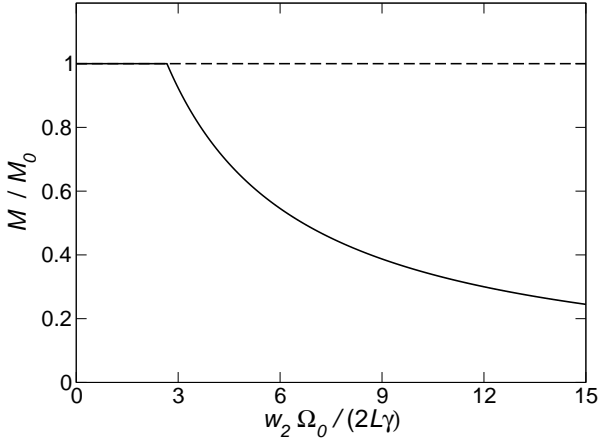}}
\end{picture}
\caption{
Magnetization $M$ of the excitonic insulator 
vs intervalley exchange interaction 
$(\Omega_0 w_2 / 2L) /\gamma$ (solid line).
The dashed line is the magnetization $M_0$ of the noninteracting phase,
which is metallic.
Here $T=0$ and
$W_0 = 1.33\gamma$. 
\label{M}}
\end{figure}

The condensation of excitons ($\Delta \neq 0$) decreases the magnitude of
$M$ with respect to $M_0$, 
making the excitonic insulator a weak paramagnet as opposed
to the normal phase, according to 
\begin{equation}
M = M_0 \frac{4\gamma}{W_0 + \Omega_0w_2/(2L) },
\end{equation}
which is valid for $W_0 + \Omega_0w_w/(2L)> 4\gamma$.
The reason is that, as electron-hole pairs 
spontaneously form, electron states in the conduction band acquire a fractional 
occupation ($v_k^2>0$) as well as hole states in the valence band ($u_k^2<1$), 
as shown in Fig.~\ref{f:clock}(b),
thus $M< M_0$. 
The magnetizations of noninteracting (dashed line) and EI (solid curve) phases 
are compared in Fig.~\ref{M}
as a function of intervalley exchange interaction.
The lower bound of the magnetization 
is $M=0$, corresponding
to the maximum exciton density allowed, the valence band being
half-empty and the conduction
band half-filled. 

The quenching of magnetization causes an increase
of the kinetic energy of the EI ground state that is  
compensated by the energy gain due to the condensation of excitons.
In fact, it is easy to show that
the kinetic energy increases linearly with the field 
whereas the interaction energy decreases
quadratically to the leading order.
This may be also seen by computing
the work $U$ done by the external field $B$ to magnetize the tube, 
$U = \int\! B\,dM$. The difference $\Delta U$ between the magnetization
work in the noninteracting and EI phases
is the condensation energy of the excitonic insulator,
\begin{equation}
\Delta U = \mu_0B^2\frac{ALe}{2ch}
\left[1-\frac{4\gamma}{W_0 + \Omega_0w_2/(2L) }\right],
\end{equation}
whose expression is valid for $W_0 + \Omega_0w_w/(2L)> 4\gamma$.

A finite amount of this condensation energy is released when 
creating a quasiparticle,
i.e., an electron (hole) occupying a definite level $k$
with unit probability. This is achieved by 
breaking the electron-hole pair of the condensate having amplitude $u_k$
for the conduction level $k$ being empty and $v_k$
being occupied. 

The quasiparticle energy $E(k)$ is defined as
the change in the ground-state energy $E_0(N)$ of the
system with $N$ electrons when adding / removing one 
particle,\cite{Abrikosov1975}
$E(k)=E_0(N \pm 1)-E_0(N)$. 
Therefore, $E(k)$
takes into account the released condensation energy 
per broken pair
in terms of the excitonic gap $\Delta$ [Eq.~\eqref{eq:Eqp}]. 
Similarly, the magnetic dipole per
particle $\mu = -\partial E(k)/\partial B$ accounts for the increase of the 
ground-state magnetization when annihilating an exciton of the condensate.
This explain the enhancement of $\mu$ with respect to the noninteracting 
value $\mu_0$
[Eq.~\eqref{eq:mu}].

The same rationale for the enhancement of the magnetic moment per
quasiparticle is valid for semiconducting tubes ($\nu=\pm 1$), as the
above argument may be applied separately to each valley in reciprocal space.
The difference with respect to metallic tubes is that the EI ground state
magnetization is now zero, since the bands in the two valleys exhibit
opposite chiralities.

\section{Relation to experiments}\label{s:exp}

The excitonic insulator phase of carbon nanotubes
has two experimental signatures of
genuine many-body origin that may be accessed by current
experiments. 

\begin{table*}
\begin{ruledtabular}
    \begin{tabular}{ c c c c c c c c }
  Reference                                       
& $\mu$ measured            
& $R$ inferred      
& $R$ measured
& $E^{\text{SO}}_g$ measured 
& $E^{\text{SO}}_g$ expected  
& $w_2$ estimated   
& $R$ predicted here \\

&  (meV/T)
& (nm)
& (nm)
& ($\mu$eV)
& ($\mu$eV)
& (eV)
& (nm) \\
    Kuemmeth 2008\cite{Kuemmeth2008}     &  1.55 &  3.50 & n.a. & 370  & 110 & 1208 & 1.73  \\
   Jespersen 2011\cite{Jespersen2011}    &  0.63 &  1.45 & n.a. & 150  & 168 &  n.a. & 1.45  \\
 Jespersen PRL 2011\cite{Jespersen2011b} &  0.87 &  2.65 & n.a. & 200  & 146 &  914.4 & 2.02  \\
 Churchill 2009\cite{Churchill2009}      &  0.33 &  0.75 & n.a. & 170  & 520 &  n.a. & 0.75  \\
 Jhang 2010\cite{Jhang2010}              &  n.a. &  n.a. & 0.75 & 2500 & 520 & 1250 & 0.67 \\
 Steele 2012\cite{Steele2013} device 1   &  1.60 &  3.60 & 1.50 & 3400 & 260 & 1242 & 0.57 \\
 Steele 2012\cite{Steele2013} device 2   &  1.50 &  3.40 & n.a. & 1500 & 116 & 1173 & 0.84 \\
 Steele 2012\cite{Steele2013} device 3   &  0.90 &  2.05 & n.a. & 1700 & 190 &
707.4 & 0.62 \\
   \end{tabular}
\caption {Comparison between predictions for the CNT excitonic 
insulator and available
experimental data. The meaning of different entries is as follows:
$\mu$ is the orbital magnetic moment measured through single-electron tunneling
spectroscopy. The inferred radius $R$ is obtained from the measured value of 
$\mu$ through the semiclassical formula $\mu_0 = e v_F R /2c$.
The measured values of $R$ are obtained from AFM experiments.
The spin-orbit energy splitting $E^{\text{SO}}_g$ is measured
through single-electron tunneling
spectroscopy in the conduction band. 
The expected value of $E^{\text{SO}}_g$ is obtained from 
the inferred value of $R$  
(or from the measured value when available) according to the 
noninteracting theory of Ref.~\onlinecite{Steele2013} (cf.~Table S1).
The intervalley contact interaction $w_2$ 
and the predicted value of $R$ in the EI phase are evaluated through
the procedure explained in the main text.
}
\label{comparison}
\end{ruledtabular}
\end{table*}

\begin{figure}
\setlength{\unitlength}{1 cm}
\begin{picture}(8.5,5.8)
\put(0.3,0.0){\epsfig{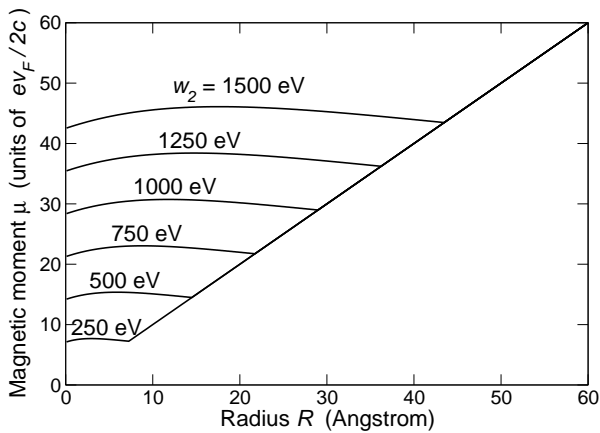}}
\end{picture}
\caption{Quasiparticle magnetic moment $\mu$
of a nominally metallic tube
vs tube radius $R$,
for different values of the intervalley contact
interaction $w_2$.
Here $T=0$ and $W_0 = 1.33 \gamma$. 
\label{f:magmom}
}
\end{figure}

The first fingerprint is the enhancement of the electron
magnetic moment $\mu$ measured by tunneling spectroscopy\cite{Minot2004} 
with respect to the semiclassical value 
$\mu_0$, which was illustrated in Sec.~\ref{s:anomalous}.
According to formula \eqref{eq:mu},
$\mu$ displays a peculiar dependence on the tube radius $R$,
in contrast to the simple linear dependence of $\mu_0$.
As shown in Fig.~\ref{f:magmom} for nominally metallic tubes, 
$\mu$ is almost independent from $R$ for small values of the radius, 
exhibiting a plateau whose extension increases 
with the intervalley contact interaction $w_2$. 
As the radius reaches the critical value of the transition from the
EI to the normal phase, $\mu$ regains the familiar linear dependence from $R$. 
An important practical consequence of this prediction is that
the radius
of a small-diameter tube may not be inferred from a measure of $\mu$, 
since only for $R$ larger than the critical size the
semiclassical formula 
$\mu_0 = e v_F R /2c$ holds.

The second fingerprint is the increase of the quasiparticle
gap of the EI phase with respect to the gap
$E_g$ of the noninteracting ground state. This occurs
through the excitonic term $\Delta$ that adds quadratically to $E_g$,
hence the EI gap may be written as
\begin{equation}
2\sqrt{(E_g/2)^2+\left|\Delta\right|^2}.
\label{eq:gapgeneric}
\end{equation}
Throughout this paper we have identified $E_g$ as the gap due solely 
to tube chirality and/or axial magnetic field,
$E_g/2\equiv \gamma
\left|k_{\nu}(0)\right|$.
Nevertheless, the result \eqref{eq:gapgeneric} is generic to
any energy gap that originates from the effective
displacement of Dirac cone apexes with respect to allowed wave vectors
in the Brillouin zone, 
including the small mass terms due to tube curvature, strain, 
and twists.\cite{Kane1997,Charlier2007}

\begin{figure}
\setlength{\unitlength}{1 cm}
\begin{picture}(8.5,6.5)
\put(0.3,0.0){\epsfig{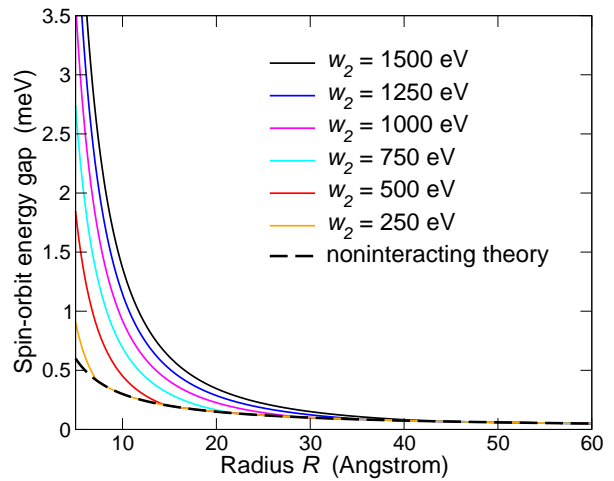}}
\end{picture}
\caption{(Color online)
Transport gap 
induced by spin-orbit interaction 
in a nominally metallic tube
vs tube radius $R$. 
The dashed line is the prediction of noninteracting theory,
$E^{\text{SO}}_g = 3 /(R \left[\text{\AA}\right])$ meV,
whereas the solid lines are the gaps in the excitonic insulator phase
for different values of the intervalley contact
interaction $w_2$.
Here $T=0$, $B=0$, and $W_0 = 1.33 \gamma$.
\label{f:SO}
}
\end{figure}

This holds also for the mass term induced by
spin-orbit interaction in nominally metallic 
tubes.\cite{Ando2000,Huertas-Hernando2006,Steele2013}
The spin-orbit correction to the noninteracting single-particles energies
\eqref{eq:energy} and \eqref{eq:energyKpri}
takes the form
\begin{equation}
\varepsilon^{\tau\sigma}_{\alpha}(k) = s_{\alpha}  \gamma
\sqrt{ k^2_{\nu=0}(n) + k^2     } ,
\label{eq:energySO} 
\end{equation}
with the transverse momentum $k_{\nu=0}(n)$ being displaced
by the spin-orbit term $\Delta_{\text{SO}}$, with phase given
by $\tau\sigma$,
\begin{equation}
k_{\nu=0}(n) =
\frac{1}{R} \left( n + \varphi 
\right) + \frac{\Delta_{\text{SO}}}{\gamma R}\tau\sigma.
\label{eq:k_xSO} 
\end{equation}
Here we ignore the rigid energy shift due to the
recently discussed Zeeman-like
term.\cite{Izumida2009,Jeong2009,Klinovaja2011,Steele2013}
Then, at zero field spin-orbit interaction opens a 
small spin-independent gap $E^{\text{SO}}_g$
between conduction and valence bands,
whose size is $E^{\text{SO}}_g = 2\Delta_{\text{SO}}/R$.

Figure \ref{f:SO} shows
the spin-orbit gap enhanced by the excitonic order (solid lines)
vs $R$. This is compared with 
the noninteracting spin-orbit gap $E^{\text{SO}}_g$, shown as
a dashed curve
(with\cite{Kuemmeth2008}
$\Delta_{\text{SO}}=1.5$ meV$\cdot$\AA).
Whereas at large $R$ the gap is inversely proportional to the radius,
consistently with the noninteracting theory, as $R$ falls
below the critical
size the gap acquires a much stronger dependence on $R$, scaling like
$1/R^2$.

Overall, the behavior of the spin-orbit gap exhibited in Fig.~\ref{f:SO}
complements that of the magnetic moment illustrated in Fig.~\ref{f:magmom},
with both observables $\mu$ and $E^{\text{SO}}_g$ 
being enhanced by an excitonic
factor that scales as $1/R$.
To derive the exact relation between $\mu$ and spin-orbit energy
splittings requires
to consider the case in which 
both spin-orbit interaction and magnetic field are present,
lifting spin and valley degeneracies of quasiparticle levels. 
This analysis will be presented elsewhere.

\subsection{Comparison with available literature}

Table \ref{comparison} lists the 
magnetic moments
and spin-orbit energy splittings reported for
single-wall CNTs close to charge neutrality at
low temperature (after Table S1
of Ref.~\onlinecite{Steele2013}). The second and third columns 
report respectively
tunneling-spectroscopy data\cite{Minot2004} for $\mu$  
and corresponding values of $R$ 
inferred through the semiclassical formula $\mu_0=ev_FR/2c$.
In all cases but two (Refs.~\onlinecite{Jespersen2011}
and \onlinecite{Churchill2009})
the inferred values exceed the actual values of $R$
by a factor two or more,   
as realistic numbers fall in a range between 0.2 and 1.5 nm.
This discrepancy is confirmed by available AFM measurements 
(fourth column), as it is the case of device 1 of Ref.~\onlinecite{Steele2013},
whose inferred radius of 3.6 nm is more than twice larger
than the measured radius of 1.5 nm.

These data are presently not understood.\cite{Laird2014}
A possible explanation 
is that the Fermi velocity $v_F$ appearing in the expression for $\mu_0$ 
has to be renormalized.\cite{Jespersen2011b} However,
a correction of a factor two can hardly be justified,  
as state-of-the-art quasiparticle calculations, able to rationalize
optical spectra, provide values of 
$v_F$ comparable to the tight-binding estimate.\cite{Choi2013}
Besides, the effective-mass corrections to $v_F$ predicted by \eqref{eq:group}
become significant only far from band edges.\cite{Jespersen2011,Jespersen2011b}

In the spirit of Sec.~\ref{s:instability}, here we use the values
of $\mu$ reported in the second column of Table \ref{comparison}
to estimate the intervalley exchange interaction parameter $w_2$.
First, from the inspection of the apparent magnitudes of $R$,
we assign all inferred values
but those of Refs.~\onlinecite{Jespersen2011} and \onlinecite{Churchill2009}
to the EI phase.
As shown in Fig.~\ref{f:magmom},
we expect that the
actual values of $R$ fall somewhere in the region
in which $\mu$ exhibits a plateau.  
Then, to extract $w_2$, we equate the experimental value of $\mu$ with that
predicted at the transition point between EI and noninteracting phase, 
being representative of the values of $\mu$ over the whole plateau. 
The outcome of this procedure is reported in the seventh column of 
Table \ref{comparison}, providing estimates for $w_2$ that vary between
700 and 1200 eV. To extract the actual tube radius
we need a second, independent set of measurements, 
which is provided by spin-orbit data. 

Table \ref{comparison}
shows available data for spin-orbit energy splittings (fifth column),
together with their maximum sizes expected by noninteracting theory
(sixth column),
according to Ref.~\onlinecite{Steele2013}.
The comparison highlights that measured data may exceed theoretical 
estimates by one order of magnitude, which has not been explained 
yet.\cite{Steele2013}
The listed experimental values have diverse origins,
i.e., they are typically energy splittings
separating excited electron states of both metallic
and semiconducting tubes, ascribed to both 
Zeeman-like and orbital-like spin-orbit terms.
Therefore, these data cannot be straightforwardly compared with 
the predictions of Fig.~\ref{f:SO}. Nevertheless, 
we use them as rough estimates of 
spin-orbit gaps in nominally metallic tubes.

Since in Table \ref{comparison}
we have now linked the samples 
to their values of $w_2$ (for the sample of Ref.~\onlinecite{Jhang2010}
we assume $w_2 = 1250$ eV), we may plot the
spin-orbit gap of each device versus $R$, similarly to the curves 
in Fig.~\ref{f:SO}. By intersecting each 
curve with the horizontal line
that identifies the measured entry of $E_g^{\text{SO}}$, we eventually 
extract the value of $R$ in the EI phase, 
which is tabulated in the last column.
This number, systematically lower than the apparent value of $R$
of column two, ranges between 0.6 and 2 nm, reasonably comparing 
with realistic values. 
This estimate provides an important consistency check of the theory.

\subsection{The Caltech experiment}

The claim of the observation of the Mott-Hubbard gap 
in nominally metallic CNTs 
by the Caltech group in 2009 
relies on the idea of using the magnetic field
to remove the noninteracting energy
contribution $E_g$ to the transport gap.\cite{Deshpande2009} 
The term $E_g$ is generically ascribed
to a small shift of the transverse wave vector  
\eqref{eq:k_xSO}, which may be fully
counteracted by a properly chosen Aharonov-Bohm phase $\varphi$
of like magnitude and
opposite sign.
At such critical value of the field
the measured transport gap was found not to vanish, even after subtracting
the energy contribution due to Coulomb blockade. Therefore,
this inherent gap
was interpreted as  
a genuine many-body effect.\cite{Deshpande2009}

Since the opening of a many-body gap at $E_g=0$
is peculiar to 
the Mott-Hubbard scenario (cf.~Table \ref{t:comparison}),
it would appear that
evidence rules out exciton condensation.
However, the presence of spin-orbit interaction,
which was overlooked in the analysis of Ref.~\onlinecite{Deshpande2009},
makes the EI scenario possible. This may be seen qualitatively 
from the spin-valley dependence of the
transverse wave vector \eqref{eq:k_xSO}: If the gap $E_g^{\text{SO}}$
in valley $\tau$ closes in the spin channel $\sigma$ for a
suitable value of $\varphi$, then  
it remains
finite in the other channel $-\sigma$, with
$E_g^{\text{SO}}=4\left|\Delta_{\text{SO}}\right|$.   
Therefore, the noninteracting gap experienced
by triplet excitons never vanishes.

Also, 
the observation of subgap neutral collective excitations reported 
in Ref.~\onlinecite{Deshpande2009} might be consistent 
with the EI scenario (cf.~Table \ref{t:comparison}).
These excitations are expected as phonon-like collective modes of the EI,
either of acoustic or optical type,
as well as spin density waves. These modes correspond to space-time
fluctuations
of the phase and magnitude of $\Delta$ or
oscillations of the spin-polarization vector
$\bm{n}$, respectively.\cite{Kohn1967,Halperin1967,Bascones2002}
The theory of EI collective excitations 
as well as the quantitative analysis of the data of 
Ref.~\onlinecite{Deshpande2009} 
are left to future work.

\section{Conclusions}\label{s:end}

In this work we have proposed that the ground state
of an undoped carbon nanotube of small diameter might
be an excitonic insulator. The condensate is made of triplet 
excitons that are stabilized by intervalley exchange interaction,
which induces antiferromagnetic spin density wave order.
The ultimate validation of this theory 
relies on the observation of 
the excitonic enhancement of both energy gap
and magnetic moment of quasiparticles, which might have already 
been measured.  

\begin{acknowledgments}
We thank Shahal Ilani, Elisa Molinari, Daniele Varsano,
Deborah Prezzi, and Ehud Altman for stimulating discussions.
This work is supported by EU-FP7 Marie Curie initial training 
network INDEX and MIUR-PRIN2012 Project MEMO. 
\end{acknowledgments}

\appendix

\section{Generic solution of the Bethe-Salpeter equation
for the lowest triplet exciton}\label{a:generictriplet}

In this Appendix we solve the Bethe-Salpeter 
equation \eqref{eq:BSE_triplet} for the lowest triplet exciton in a  
semiconducting carbon nanotube ($\nu = \pm 1$) in the presence
of an external magnetic field along the tube axis.
The method is a generalization of the procedure
explained in Sec.~\ref{s:twoband}, which we work out numerically.

The generic solution of \eqref{eq:BSE_triplet} is a bound state 
with distinct envelope functions in the two valleys,
\begin{eqnarray}
\psi_{\text{K}}(y) & = & \sin{\theta}
\sqrt{ \kappa_{\text{K}} }
\,\exp( -\kappa_{\text{K}} 
\left|y\right| ),\nonumber\\
\psi_{\text{K}'}(y) & = & \cos{\theta}
\sqrt{ \kappa_{\text{K}'} }
\, \exp( -\kappa_{\text{K}'} \left|y\right| ),
\end{eqnarray}
with $0\le \theta\le \pi/2$, whose energy is
\begin{equation}
\varepsilon_u = 2\gamma\sqrt{k_{\nu}(0)^2 - \kappa_{\text{K}}^2 } 
= 2\gamma\sqrt{k_{-\nu}(0)^2 - \kappa_{\text{K}'}^2 }.
\label{eq:sqrt_e_generic}
\end{equation}
The exciton energy $\varepsilon_u$ is smaller than the minimum between the
two
valley-dependent energy gaps 
$2\gamma\left|k_{\nu}(0)\right|$ and 
$2\gamma\left|k_{-\nu}(0)\right|$. If these
are equal (either the tube is metallic 
or the field is zero) than $\theta = \pi/4$ and one recovers
the solution \eqref{eq:BSEsolution}. 

The boundary condition at the origin \eqref{eq:boundary2}
now turns into the system of equations
\begin{subequations}
\label{eq:BSEsystem}
\begin{eqnarray}
&& \left[ \gamma\sqrt{k_{\nu}(0)^2 - \kappa_{\text{K}}^2 } 
- \gamma\left|k_{\nu}(0)\right| 
+\frac{\kappa_{\text{K}} W_0}{4} \right] \sin{\theta}
\sqrt{ \kappa_{\text{K}}   } \nonumber \\
&& = \quad -\frac{\kappa_{\text{K}} \Omega_0 w_2}{8L}\cos{\theta}
\sqrt{ \kappa_{\text{K}'} },\\
&& \left[ \gamma\sqrt{k_{-\nu}(0)^2 - \kappa_{\text{K}'}^2 } 
- \gamma\left|k_{-\nu}(0)\right| 
+\frac{\kappa_{\text{K}'} W_0}{4} \right] 
\cos{\theta}\sqrt{ \kappa_{\text{K}'} }
\nonumber \\
&& = \quad -\frac{\kappa_{\text{K}'} \Omega_0 w_2}{8L}
\sin{\theta}\sqrt{ \kappa_{\text{K}} } .
\end{eqnarray}
\end{subequations}
As the onset of the excitonic instability occurs for $\varepsilon_u = 0$,
it is easy to show that the critical value of intervalley exchange
interaction is given again by Eq.~\eqref{eq:critical},
$W_0 + \Omega_0 w_2/(2L) = 4\gamma$, which implies 
\begin{equation}
\tan{\theta}=\sqrt{\left|\frac{k_{-\nu} (0)}{k_{\nu} (0)} \right|    }.
\end{equation}

Solving system \eqref{eq:BSEsystem}
for $\kappa_{\text{K}}$ and $\kappa_{\text{K}'}$ other than zero
leads to the secular equation 
\begin{eqnarray}
&& \left[ \gamma\sqrt{k_{\nu}(0)^2 - \kappa_{\text{K}}^2 } 
- \gamma\left|k_{\nu}(0)\right| 
+\frac{\kappa_{\text{K}} W_0}{4} \right] \nonumber \\
&&\times\quad \left[ \gamma\sqrt{k_{-\nu}(0)^2 - \kappa_{\text{K}'}^2 } 
- \gamma\left|k_{-\nu}(0)\right| 
+\frac{\kappa_{\text{K}'} W_0}{4} \right]\nonumber\\
&&\quad-\quad \kappa_{\text{K}}\kappa_{\text{K}'}\left(\frac{\Omega_0 w_2}{8L}
\right)^2=0.
\label{eq:secular}
\end{eqnarray}
We obtain the root of \eqref{eq:secular} numerically,
with both $\kappa_{\text{K}}$ and $\kappa_{\text{K}'}$ 
explicited in terms of the unknown $\varepsilon_u$.
This gives the exciton energy at arbitrary 
values of energy gaps, as illustrated in Fig.~\ref{supercritical}.

\section{Generic solution of the EI gap equation}\label{a:genericgap}

In this Appendix we solve the gap
equation \eqref{eq:gap2} for a
semiconducting carbon nanotube ($\nu = \pm 1$) in the presence
of an external magnetic field applied along the tube axis.

Paralleling the strategy of Sec.~\ref{s:gap}, we assume that 
the excitonic gap in each valley is much smaller than the
corresponding noninteracting gap. Therefore, we may
expand the square roots
entering \eqref{eq:gap2} and obtain
\begin{eqnarray}
&&-\frac{\left|\Delta(\text{K}  )
\right|^2}{\gamma \left|k_{\nu}(0)\right| } \varphi^{\text{K}}(k)
= 
2\gamma \sqrt{ k^2_{\nu}(0) + k^2 }   \varphi^{\text{K}}(k) \nonumber \\
&& - \frac{1}{A} \sum_q W_0 \varphi^{\text{K}}(k+q) 
-\frac{1}{A} \sum_q
\frac{\Omega_0 w_2}{2L} \varphi^{\text{K}'}(k+q) , \nonumber \\
&&-\frac{\left|\Delta(\text{K}' )
\right|^2}{\gamma \left|k_{-\nu}(0)\right| } \varphi^{\text{K}'}(k)
= 
2\gamma \sqrt{ k^2_{-\nu}(0) + k^2 }   \varphi^{\text{K}'}(k) \nonumber \\
&& - \frac{1}{A} \sum_q W_0 \varphi^{\text{K}'}(k+q) 
-\frac{1}{A} \sum_q
\frac{\Omega_0 w_2}{2L} \varphi^{\text{K}}(k+q).
\label{eq:gap4}
\end{eqnarray}
This is identical to the Bethe-Salpeter 
equation \eqref{eq:BSE_triplet} for a triplet 
exciton of energy
\begin{equation}
\varepsilon_u = -\frac{\left|\Delta(\text{K}  )
\right|^2}{\gamma \left|k_{\nu}(0)\right| }
= -\frac{\left|\Delta(\text{K}' )
\right|^2}{\gamma \left|k_{-\nu}(0)\right| }.
\label{eq:DeltaKKpr}
\end{equation}
We exploit this identity to
obtain the excitonic gap using the numerical results derived
by the method explained in Appendix \ref{a:generictriplet}.


%

\end{document}